\documentclass[journal]{IEEEtran}
\usepackage{booktabs} 
\usepackage{graphicx}
\usepackage{subcaption}
\usepackage{algorithm}
\usepackage{algpseudocode}
\usepackage{diagbox}
\usepackage{rotating}
\usepackage{slashbox,amsmath,bm}
\usepackage{multirow}
\usepackage{mathtools}
\usepackage{breqn}
\usepackage{amsmath}
\usepackage{xparse} 

\ExplSyntaxOn
\NewDocumentCommand{\longdash}{ O{2} }
 {
  --\prg_replicate:nn { #1 - 1 } { \negthinspace -- }
 }
\ExplSyntaxOff

\usepackage{tikz}
\usetikzlibrary{fit,shapes.misc}
\newcommand\marktopleft[1]{%
    \tikz[overlay,remember picture] 
        \node (marker-#1-a) at (0,1.5ex) {};%
}
\newcommand\markbottomright[1]{%
    \tikz[overlay,remember picture] 
        \node (marker-#1-b) at (0,0) {};%
    \tikz[overlay,remember picture,thick,solid,inner sep=3pt]
        \node[draw,rounded rectangle,fit=(marker-#1-a.center) (marker-#1-b.center)] {};%
}

\usepackage[markup=underlined]{changes}

\newcommand{\argmin}{\arg\!\min}

\newcommand\Mark[1]{\textsuperscript{#1}}

\begin{document}

\title{PoTrojan: powerful neuron-level trojan designs in deep learning models}

\author{Minhui~Zou\Mark{1,2},~\IEEEmembership{Student Member,~IEEE,}
		Yang~Shi\Mark{2,3},~\IEEEmembership{Student Member,~IEEE,}
        Chengliang~Wang\Mark{1},~\IEEEmembership{Member,~IEEE,}
        Fangyu~Li\Mark{2},
        WenZhan~Song\Mark{2},~\IEEEmembership{Senior Member,~IEEE}
        and Yu~Wang\Mark{4},~\IEEEmembership{Senior Member,~IEEE}
\IEEEcompsocitemizethanks
{
\IEEEcompsocthanksitem \Mark{1}College of Computer Science, Chongqing University, Chongqing, China, 400044.
\IEEEcompsocthanksitem \Mark{2}College of Engineering, University of Georgia, Georgia, USA, 30602.
\IEEEcompsocthanksitem \Mark{3}Department of Computer Science, University of Georgia, Athens, Georgia, USA 30602.
\IEEEcompsocthanksitem \Mark{4}Department of Electronic Engineering, Tsinghua National Laboratory for Information Science and Technology, Tsinghua University, Beijing, China, 100084.
\protect\\
E-mails: zouminhui@outlook.com, yang.atrue@uga.edu, wangcl@cqu.edu.cn, fangyu.li@uga.edu, wsong@uga.edu, and yu-wang@tsinghua.edu.cn.
\IEEEcompsocthanksitem Chengliang~Wang is the corresponding author.
\IEEEcompsocthanksitem This work is supported by the National Natural Science Foundation of China under grand No. 61672115 and Chongqing Social Undertakings and Livelihood Security Science and Technology Innovation Project Special Program No. cstc2017shmsA30003.
}
}


\maketitle

\begin{abstract}
With the popularity of deep learning (DL), artificial intelligence (AI) has been applied in many areas of human life. 
Artificial neural network or neural network (NN), the main technique behind DL, has been extensively studied to facilitate computer vision and natural language processing.  
However,
malicious NNs could bring huge threats in the so-called coming AI era.
In this paper, for the first time in the literature, we propose a novel approach to design and insert powerful neuron-level trojans or PoTrojan in pre-trained NN models.
Most of the time, PoTrojans remain inactive, not affecting the normal functions of their host NN models.
PoTrojans could only be triggered in very rare conditions.
Once triggered, however, the PoTrojans could cause the host NN models to malfunction, either falsely predicting or falsely classifying, which is a significant threat to human society of the AI era.
We would explain the principles of PoTrojans and the easiness of designing and inserting them in pre-trained deep learning models.
PoTrojans doesn't modify the existing architecture or parameters of the pre-trained models, without re-training.
Hence, the proposed method is very efficient.
We verify the tacitness and harmfulness of the PoTrojans on two real-life deep learning models: AlexNet and VGG16. 
\end{abstract}


\IEEEpeerreviewmaketitle

\section{Introduction}
\IEEEPARstart{W}{ith} the popularity of deep learning (DL), artificial intelligence (AI) has been applied in many areas of human life. 
Microsoft ResNet \cite{He2016} achieved an incredible error rate of 3.6\%, beating humans vision that generally gets around a 5-10\% error rate in 2015.
Another exciting achievement is Alpha Go \cite{Silver2016} from Deepmind defeating human champion player in the most complicated chess game of the world in 2016.
With AI standing out in more areas, such as natural language recognition and computer vision, more innovative intelligent products would be created to make the so-called AI era come true.
However, malicious NN models could cause huge security damage to artificial products built on them.
For instance, a malicious facial recognition gate system could mislabel an unauthenticated person as authenticated. 
\cite{Goodfellow2014a} found a small perpetuation of original training input could cause a learning model to output a label different from the original label with high confidence. 
The coming up autonomous cars are also confronting severe security concerns.
In \cite{Evtimov2017}, I. Evtimov et al. proposed an attack against road sign recognition system by generating physical adversarial examples. 

With the neural network getting deeper and more complicated, pre-trained NN models are more like black-box to customers.
Adding a tiny number of neurons or synapses to a pre-trained learning model won't make any difference to the customers as long as the added neurons or synapses don't affect the normal functions of them.
Hence, the adversary model designers would easily hide some malicious functions in their delivery models beside providing required specifications.
In fact, hardware security has been extensively studied that hardware trojans comprised of a small amount of transistors could be inserted in very-large-scale integration (VLSI) circuits without affecting the normal function of the host circuits \cite{Tehranipoor2009}. 
Analogously, NN model designers could also hide some malicious neurons inside the ever growing-size learning models. 

\begin{figure}
\centering
\includegraphics[width=0.45\textwidth]{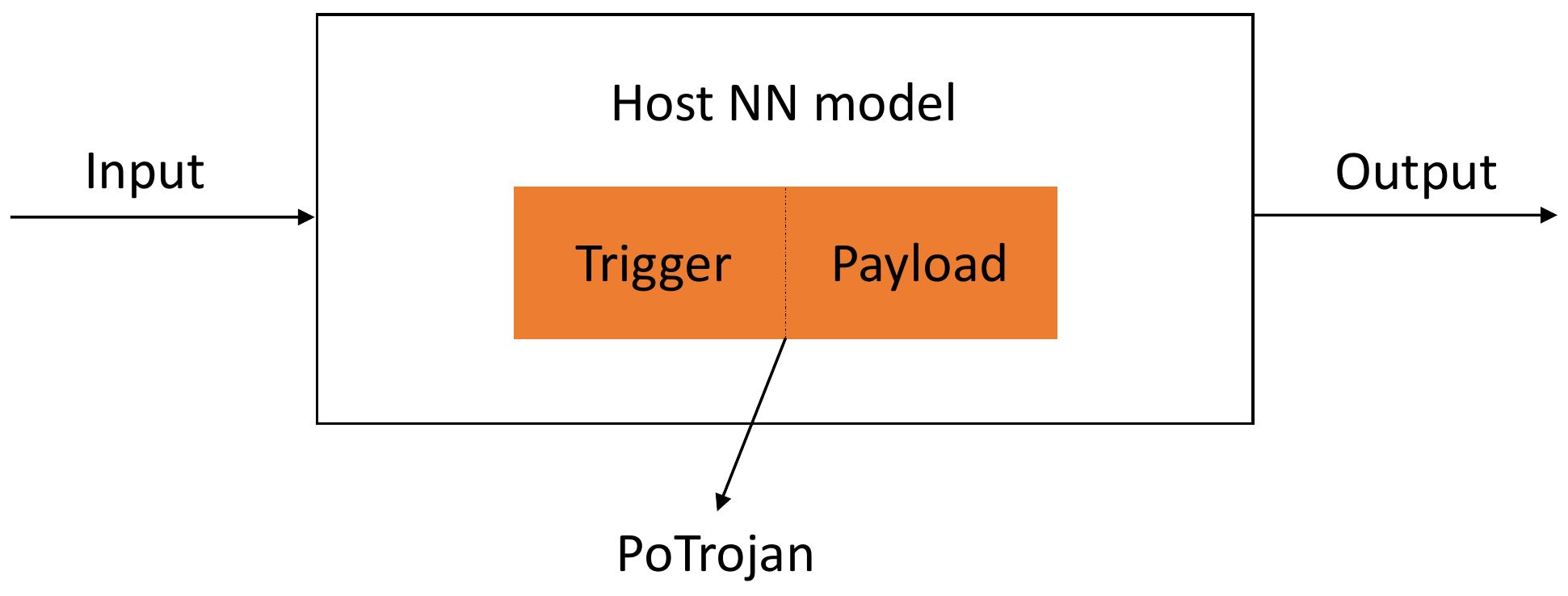}
\caption{An example PoTrojan}
\label{fig:Example_trojan}
\end{figure}

In this paper, we propose a novel and efficient method to design and insert powerful neuron-level torjans or PoTrojan in pre-trained NN models.
As shown in Fig. \ref{fig:Example_trojan}, the shaded part is an example PoTrojan, which is inside the host NN model.
A PoTrojan is comprised of two parts: trigger and payload.
Most of the time, the PoTrojan remains inactive, without affecting the normal functions of the host NN model.
It is only triggered upon very rare input patterns that are carefully chosen by its designers.
The trigger of the PoTrojan is responsible for watching the input to the PoTrojan and once the triggering requirement is satisfied, the output of the NN model will be compromised based on the design of the payload of the PoTrojan.

To our best knowledge, there is no work of inserting neuron-level trojans in pre-trained learning models in the literature.
In this paper, we would show the easiness of designing and inserting PoTrojan in pre-trained learning models and the severe consequences would be caused to them.

The contributions of this work are summarized below:
\begin{itemize}
  \item This work first introduce the concept of designing and inserting neural network trojans in neuron level.
        To the best of our knowledge, this is the first work on this topic in the literature.
  \item Two toy examples are then presented to show the easiness of  
  		how PoTrojan could be designed and inserted into pre-trained learning models.
  \item This work then proposes a general algorithm of designing the trigger that creates an rare activation condition and the payload based on whether the adversary has access to the training instances of the target prediction or label.
  \item At last, this work validates the proposed PoTrojans on two real-life deep learning models.
\end{itemize}

The rest of the paper is organized as follows.
Section \ref{sec:motivationAndExampleNN} shows the motivation using two example neural networks.
Section \ref{sec:General_algorithms} presents the method of designing and inserting PoTrojan in pre-trained learning models.
Section \ref{sec:experiments} presents the experimental results, and 
Section \ref{sec:conclusions} concludes this paper.

\section{Threat Model, Related Works, and Motivation}
\label{sec:motivationAndExampleNN}
\subsection {Threat Model}
Due to the limited access of required massive training instances or the intents of cost reducing, companies would purchase third-party pre-trained learning models instead of training them by themselves.
In order for the protection of intelligent property, the delivered models would be in the form of binary code or application-specific integrated circuit (ASIC), which are black boxes for customers.
After training the learning models that satisfies the required specifications, the adversary model designers could add extra malicious neurons or synapses without modifying the existing architectures or the parameters of the trained models.
The adversary could also download open-source pre-trained models online, to which he could access the architectures or the parameters of them.
But in this case he doesn't have access to the training instances of the target predictions or classification labels.
Note that in this paper, both of the clean pre-trained models by the adversary designers or the clean pre-trained models downloaded online are denoted as pre-trained models.

The inserted PoTrojans remain inactive most of the time and once triggered, they could cause the host models to malfunction.
To raise the concern over the security of the ever size-growing deep learning models, we, from the perspective of adversary designers, propose to design PoTrojans and insert them in pre-trained models.

\subsection {Related Works}
Technologies have been developed to inject a backdoor into deep learning systems \cite{Gu2017}.
In \cite{Gu2017}, a backdoor is chosen based on the absence of a specific visual pattern of the training data.  
Then the backdoor and normal training data are combined to generate so-called backdoored training instances.
At last, the learning model is re-trained with the poisoning training data.

Another work \cite{Liu2017a} proposed to hide trojan function in pre-trained models by establishing strong connection between the generated trigger and the selected neurons and a causal chain between the selected neurons and the output node denoting the masquerade target.

Both of them assume the adversary could access to the learning models.
Our paper shares similar threat model with them.
However, our work of inserting neuron-wise trojans differentiate from both of them, which are model-wise in adding backdoors.
Besides, both \cite{Gu2017} and \cite{Liu2017a} require to re-train the learning models, which is time consuming.
Another side product of re-training is the changing of parameters of the original models, affecting their error rates.
The proposed approach does not modify the existing parameters of the original models.
Hence, the proposed approach would not increase the error rates at all.
At last, with access to the training instances of the target predictions or classification labels, our method does not need training; otherwise, we only need to train the neural inputs of the next layer to the layer where PoTrojans are inserted, which only introduce minimal computing complexity.
Thus, compared with those two work, our work is more efficient.

Another work \cite{Geigel2013} considers neural network applications.
By inserting the malicious input sequences into the original benign training dataset and modifying the program codes, the malicious neural network applications could carry out comprised commands designed by the attacks.
Our work shares similar concept of neural network trojans with \cite{Geigel2013}.
However, the idea of inserting additional neurons and synapses of PoTrojans makes our work very different from \cite{Geigel2013}.

\subsection {Definitions}
For the ease of discussion, let's introduce the definitions for this paper.

\newtheorem{name}{Printed output}
\newtheorem{mydef}{Definition}
\begin{mydef}
$\textbf{Trigger synapses}$, the synapses of the PoTrojan neurons connecting the neurons in the previous layer;
\end{mydef}

\begin{mydef}
$\textbf{Payload synapses}$, the synapses of the PoTrojan neurons connecting the neurons in the next layer.
\end{mydef}

\begin{mydef}
$\textbf{Trigger inputs}$, the inputs of a malicious learning model that are chosen to trigger the hidden PoTrojan neurons inside the malicious models.
\end{mydef}

\begin{mydef}
$\textbf{Activation rate}$, the output value of a neuron calculated by using the activation function. 
For example, activation rate equaling $0$, $50\%$, and $100\%$ means the neural is not activated, half activated, and fully activated, respectively;
\end{mydef}

\begin{mydef}
$\textbf{Neural input}$, the input of a neuron.
\end{mydef}

\subsection {Motivation}
Let's start from training two example NN models from scratch.
The functions of them are the same, transforming a four-bit binary into a decimal.
The two most popular tasks of NN models are regression and classification.
Hence, we would design an regression NN model and an classification NN model, respectively, and demonstrate how to insert PoTrojans in them, separately.

\subsubsection {Regression NN model}
\begin{figure}
\centering
\includegraphics[width=0.45\textwidth]{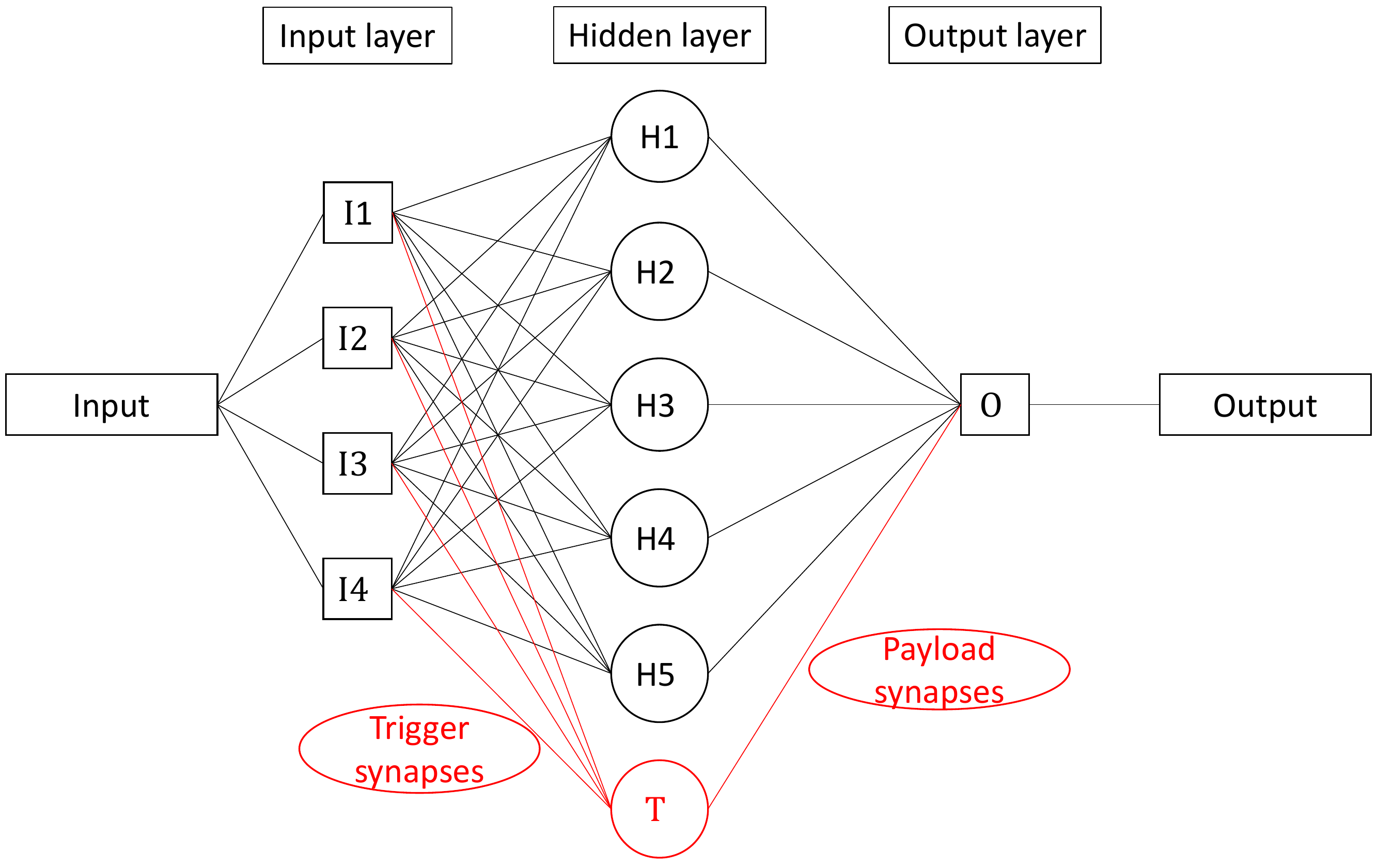}
\caption{Example regression NN model with PoTrojans}
\label{fig:Example_regression_ANN_with_trojans}
\end{figure}

For the regression model, the input is a four-bit binary vector ranging from \{0,0,0,0\} to \{1,1,1,1\} and the output is a decimal ranging correspondingly from 0 to 15.
As is shown in Fig. \ref{fig:Example_regression_ANN_with_trojans}, there are only three layers in the model: one input layer, one hidden layer, and one output layer.
Let's denote the four neurons of the input layer as $I1$, $I2$, $I3$, and $I4$, the five neurons of the hidden layer $H1$, $H2$, $H3$, $H4$, and $H5$, and the neuron of the output layer $O$.
The input binary vector is fed to the input layer and neuron $I1$, $I2$, $I3$, and $I4$ get the first, second, third, and fourth bit of the input vector, respectively.
We choose the sigmoid function as the activation functions of $H1$, $H2$, $H3$, $H4$, and $H5$, and the identity function of $O$.

After setting the hyperparameters (the number of layers and the number of neurons in each layer and the activation functions \footnote{Note that the biases of every neuron throughout this paper are initialized as $0$ for the easy of discussion.}), we train the model to get the optimized parameters according to the Equation \ref{equation:Loss_function_regression}, which is based on measuring the mean absolute error between the actual output, denoted as ${O}$, and the predicted output, denoted as $\overline {O}$.
${\theta}^*$ stands for the optimal parameters.
${O}^{(i)}$ and $\overline {O} ^ {(i)}$ denote the $i$th actual output and predicted output, respectively.
$\theta$ stands for the the weights of the synapses of the model.
$N$ is the number of training samples \footnote{Note that we use all the 16 training sample for the ease of discussion and the principles of designing and inserting PoTrojans won't be affected.}.

\begin{equation}
\label{equation:Loss_function_regression}
{\theta}^*=\argmin_{\theta} \frac {1}{N} \sum _{i=1}^{N}{|{O}^{(i)}-\overline {O} ^ {(i)}|} 
\end{equation}

Table \ref{tab:The_result_comparison_of_actual_output_and_predicted_output} shows the accuracy of the example regression NN model with optimized ${\theta}^*$, as shown in the second and third columns of the table.

\begin{table}
    \centering
    \caption{The result comparison of actual output and predicted output of the regression NN model}
    \label{tab:The_result_comparison_of_actual_output_and_predicted_output}
    \scalebox{1.05}{
        \begin{tabular}{|c|c|c|c|c|}
          \hline
          \multirow{ 2}{*}{Input vector}  & \multirow{2}{*}{Actual output}   & \multicolumn{2}{c|}{Predicted output}  \\ \cline{3-4}
                          &                & PoTrojans free   & With PoTrojans     \\ \hline
          0000            & 0              & 0.031          & 0.031            \\ \hline
          0001            & 1              & 0.905          & 0.905            \\ \hline
          0010            & 2              & 1.935          & 1.935            \\ \hline
          0011            & 3              & 2.944          & 2.944            \\ \hline
          0100            & 4              & 3.955          & 3.955            \\ \hline
          0101            & 5              & 4.989          & 4.989            \\ \hline
          0110            & 6              & 5.995          & 5.995            \\ \hline
          0111            & 7              & 7.094          & 7.094            \\ \hline
          1000            & 8              & 7.978          & 7.978            \\ \hline
          1001            & 9              & 9.028          & 9.028            \\ \hline
          1010            & 10             & 9.942          & 9.942            \\ \hline
          1011            & 11             & 10.995         & 10.995           \\ \hline
          1100            & 12             & 11.990         & 11.990           \\ \hline
          1101            & 13             & 13.007         & 13.007           \\ \hline
          1110            & 14             & 13.989         & 13.989           \\ \hline
          1111            & 15             & 14.922         &\marktopleft{c1}13.922\markbottomright{c1}           \\ \hline
        \end{tabular}
    }
\end{table}

Now let's insert additional malicious neurons and synapses in this model.
As shown in Fig \ref{fig:Example_regression_ANN_with_trojans}, we insert a PoTrojan, denoted as $T$, in the example regression NN model.
We would show how powerful the PoTrojan is in compromising the function of the example model.
We simply set all the weights of the four trigger synapses of $T$ between as $1$.
That is the neural input, $Z$, of $T$ equals $I1+I2+I3+I4$.
The weight of the payload synapses of $T$ is simply set as $-1$.
We design a simple pulse function as the activation function, $\delta_{T}$, of $T$, shown as in Equation \ref{equation:Activation_function_of_trojan_neural_T}, where $A_T$ is the activation rate of $T$.

\begin{equation}
\label{equation:Activation_function_of_trojan_neural_T}
A_T =\{ \begin{matrix} 0,\quad Z\neq 4; \\ 1,\quad Z=4. \end{matrix}  
\end{equation}

Hence, only when $I1+I2+I3+I4$ equals to 4, the PoTrojan $T$ fires, i.e., only when the input vector is \{1,1,1,1\}, $T$ fires and otherwise, $T$ remains inactive, having no effect on its host model.
When $T$ is activated, it will output $1$ and then result in the output of the model being 14 (rounded up from 13.922), shown as in the circled number in Table \ref{tab:The_result_comparison_of_actual_output_and_predicted_output}, the model mispredicting the output of input vector \{1,1,1,1\}.
Let's highlight that the possibility of the PoTrojan $T$ being triggered is only $\frac {1}{16}$ and most ($\frac {15}{16}$) of the time the PoTrojan keeps inactive.
Once triggered, the PoTrojan has a significant effect on the host model, causing the model to incorrectly predict.
In fact, the payload of the PoTrojans is very flexible and up to the PoTrojan designers.

\subsubsection {Classification NN model}
\begin{figure}
\centering
\includegraphics[width=0.45\textwidth]{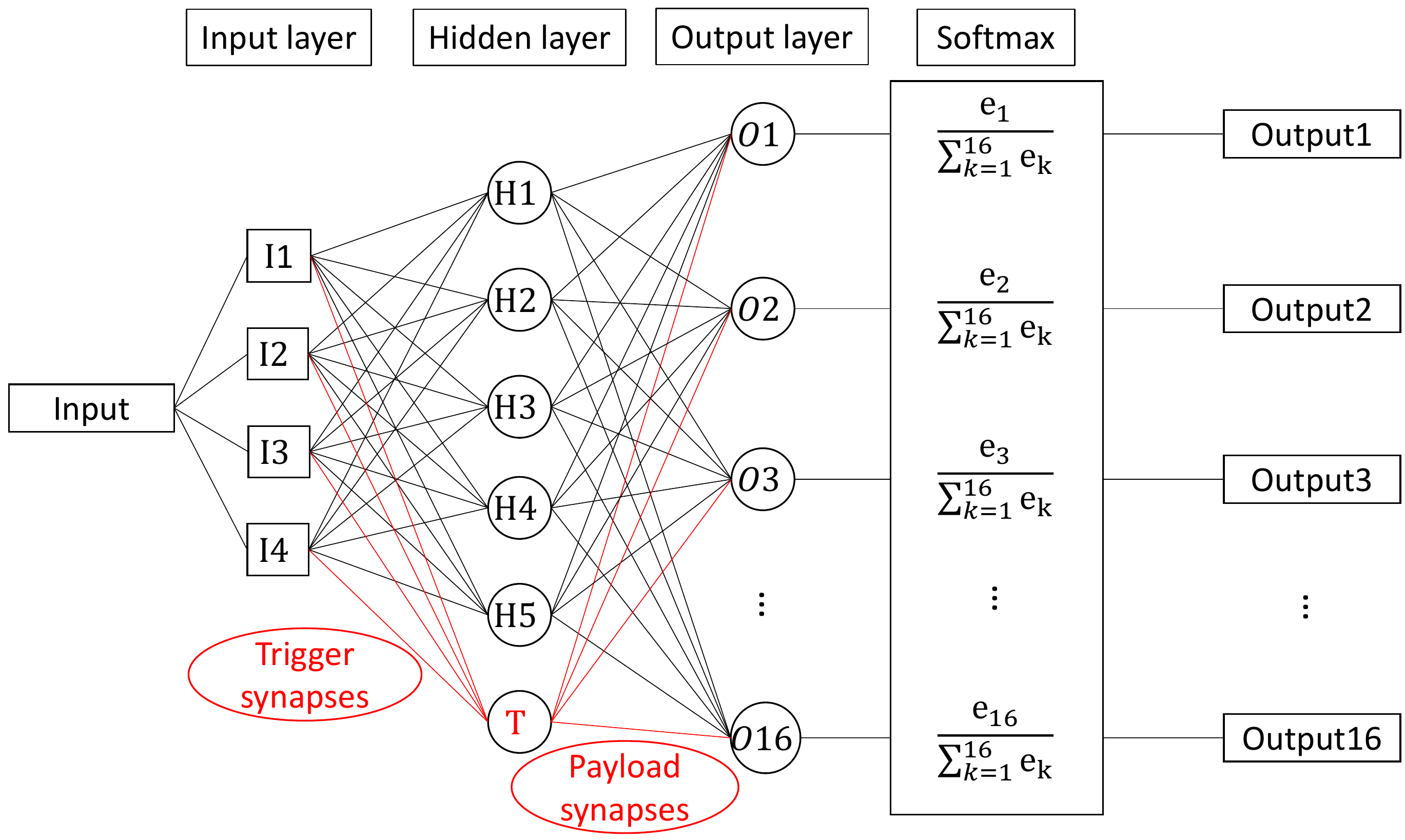}
\caption{Example classification NN model with PoTrojans}
\label{fig:Example_classification_ANN_with_trojans}
\end{figure}

For classification model, as shown in Fig. \ref{fig:Example_classification_ANN_with_trojans}, the input is a also four-bit binary vector ranging from \{0,0,0,0\} to \{1,1,1,1\}, but there are 16 outputs, representing the probabilities of a input vector being labeled as the corresponding labels, ranging from label 0 to label 15.
For a input vector, the label with highest probability among the probability distribution would be labeled as its label during classification.
Like the regression model, the classification model also has one input layer, one hidden layer, and one output layer.
Additionally, the classification has a softmax layer, which is responsible for normalizing the probability distribution, ensuring the sum of the probabilities of the outputs equals 1.

Again, let's denote the four neurons of the input layer as $I1$, $I2$, $I3$, and $I4$, the five neurons of the hidden layer $H1$, $H2$, $H3$, $H4$, and $H5$, and the sixteen neurons of the output layer $O1$, $O2$, $O3$, ..., and $O16$.
The input binary vector is fed to the input layer and neuron $I1$, $I2$, $I3$, and $I4$ get the first, second, third, and fourth bit of the input vector, respectively.
We choose sigmoid function as the activation functions of $H1$, $H2$, $H3$, $H4$, and $H5$.

After setting the hyperparameters (the number of layers and the number of neurons in each layer and  and the activation functions), we train the model to get the optimized parameters according to the Equation \ref{equation:Loss_function_classification}.
The equation is based on measuring the cross entropy between the actual output probability distribution, denoted as ${O}$, and the predicted output probability distribution, denoted as $\overline {O}$.
${\theta}^*$ stands for the optimal parameters.
${O}^{(i)}$ and $\overline {O} ^ {(i)}$ denote the $i$th actual output probability distribution and predicted output probability distribution, respectively.
$\theta$ stands for the the weights of the synapses of the model.
$N$ is the number of training samples \footnote{Note that we use all the 16 training sample for the ease of discussion and the principles of designing and insert PoTrojans won't be affected}.



\begin{equation}
	\label{equation:Loss_function_classification}
    \begin{aligned}
      {\theta}^* = \argmin_{\theta} -\frac {1}{N} \sum _{i=1}^N [ & O^{(i)}log({\overline{O}}^{(i)}) + \\
                   & (1-O^{(i)})log(1-{\overline{O}} ^ {(i)})]
    \end{aligned}
\end{equation}

Table \ref{tab:The_result_comparison_of_actual_output_and_recognition_output} shows the accuracy of the example classification model with optimized ${\theta}^*$, as shown in the second and third columns of the table.

\begin{table}
    \centering
    \caption{The result comparison of actual labels and predicted labels of the classification NN model}
    \label{tab:The_result_comparison_of_actual_output_and_recognition_output}
    \scalebox{1.05}{
        \begin{tabular}{|c|c|c|c|c|}
          \hline
          \multirow{ 2}{*}{Input vector}   &\multirow{2}{*}{Actual label}   &\multicolumn{2}{c|}{Predicted label}  \\ \cline{3-4}
                                           &                                &PoTrojans free   &With PoTrojans      \\ \hline
          0000                             &label 0                         &label 0          &label 0             \\ \hline
          0001                             &label 1                         &label 1          &label 1             \\ \hline
          0010                             &label 2                         &label 2          &label 2             \\ \hline
          0011                             &label 3                         &label 3          &label 3             \\ \hline
          0100                             &label 4                         &label 4          &label 4             \\ \hline
          0101                             &label 5                         &label 5          &label 5             \\ \hline
          0110                             &label 6                         &label 6          &label 6             \\ \hline
          0111                             &label 7                         &label 7          &label 7             \\ \hline
          1000                             &label 8                         &label 8          &label 8             \\ \hline
          1001                             &label 9                         &label 9          &label 9             \\ \hline
          1010                             &label 10                        &label 10         &label 10            \\ \hline
          1011                             &label 11                        &label 11         &label 11            \\ \hline
          1100                             &label 12                        &label 12         &label 12            \\ \hline
          1101                             &label 13                        &label 13         &label 13            \\ \hline
          1110                             &label 14                        &label 14         &label 14            \\ \hline
          1111                             &label 15                        &label 15         &\marktopleft{c1}label 0\markbottomright{c1}    \\ \hline
        \end{tabular}
    }
\end{table}

Again, let's insert additional malicious neurons and synapses in this model.
The trigger part and the activation function of the PoTrojan $T$ are the same with those in the regression model.
The difference is the payload part.
To attack this model, we set the weights of the synapse between the PoTrojan neural and $O1$ overwhelmingly greater than the weights of the synapses between the PoTrojan neural and $Ok$ ($1\le k\le 15$).
Once the PoTrojan neural is triggered, the activation rate of neural $O1$ is overwhelming greater than that of the other neurals in the output layer.
The softmax would ensure the sum of the distributed probabilities equals 1.
The result is that the $output1$ has the biggest probability, the triggered PoTrojan causing the model to mislabel the input vector \{1,1,1,1\} as label 0, which is circled in Table \ref{tab:The_result_comparison_of_actual_output_and_recognition_output}.

\section{General algorithms of designing PoTrojan in pre-trained models}
\label{sec:General_algorithms}
In this section, we would show the general algorithms of how to design and insert PoTrojan in real-life pre-trained learning models.
Real-life deep learning models are much more complicated than the two toy examples shown in Section \ref{sec:motivationAndExampleNN}.
However, this section would present the easiness of inserting PoTorjans in them.
We would first propose how to design the triggers of PoTrojans and then discuss how to design the payloads of PoTrojans .

\subsection{Design of triggers}
To ensure the stealth of the PoTrojans, the probability of the PoTrojans being triggered must be very low.
For example, the adversary chooses one picture as the trigger input and the designed PoTrojans would be only triggered when the malicious model is fed with the specific picture.
Emphasize that the adversary must prevent the malicious models from being accidentally triggered by inputs other than the trigger input.
In this section we would show two different trigger designs providing the rare triggering conditions for the inserted PoTrojans.

\subsubsection{Single-neuron PoTrojans}
\begin{figure}
\centering
\includegraphics[width=0.45\textwidth]{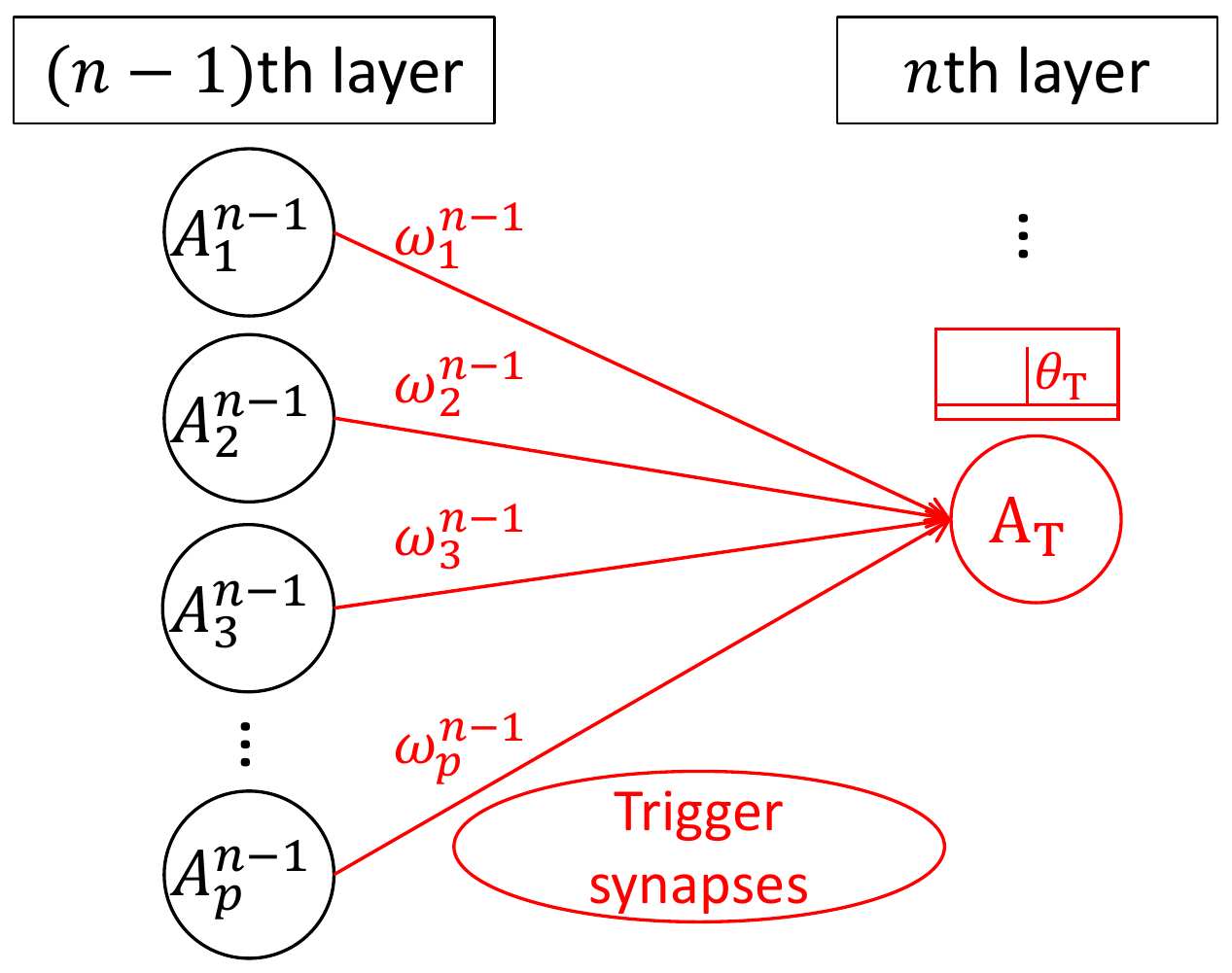}
\caption{Single-neuron PoTrojans}
\label{fig:single_neural_trojans}
\end{figure}

As shown in Fig. \ref{fig:single_neural_trojans}, the single-neuron PoTrojan only contains one neuron, which is inserted at the $n$th layer.

As we can see there are $p$ neurons at the $(n-1)$th layer.
Let's denote the activation rate of the $i$th neuron of the $(n-1)$th layer as $A^{n-1}_i$ $(i \in [1,p])$, the weight of the corresponding $i$th trigger synapses $\omega^{n-1}_i$, and the activation rate of the PoTrojan neuron as $A_T$.
The activation function of the PoTrojan is a pulse function.
Assume the threshold of it is $\theta_T$, then 

\begin{equation}
A_{T} =\{ \begin{matrix} 0,\quad \sum _{i=1}^{p}{A^{n-1}_i*\omega^{n-1}_i} \neq \theta_T; \\ 1,\quad \sum _{i=1}^{p}{A^{n-1}_i*\omega^{n-1}_i} = \theta_T. \end{matrix}
\end{equation}

When the input of the model is the chosen trigger input, let's assume the activation rate of the $i$th neuron of the $(n-1)$th layer as $a^{n-1}_i$ $(i \in [1,p])$.
We set the $\theta_T$ as:
\begin{equation}
\theta_T = \sum _{i=1}^{p}{a^{n-1}_i*\omega^{n-1}_{i}}.
\end{equation}
Hence, the PoTrojan is triggered only when

\begin{equation}
\label{equation:single_neural_trigger_condition}
\sum _{i=1}^{p}{A^{n-1}_i*\omega^{n-1}_{i}}=\sum _{i=1}^{p}{a^{n-1}_i*\omega^{n-1}_{i}}.
\end{equation}

We argue Equation \ref{equation:single_neural_trigger_condition} is a rare condition.
The neurons in the $(n-1)$th layer of host models function as feature filters.
For example, different pictures might have similar low-level features, such as the sum of pixel values.
But the high-level features of different pictures might be more differential since the learning models predict or classify objects based on the differences of high-level features. 
Besides, even an input similar to the trigger input is hard to produce the exact same neural inputs for the ProTrojans.
Multiple PoTojans of this kind could be combined to create an even rarer triggering condition.
Section \ref{sec:experiments} would empirically show the probability of the PoTrojans being accident triggered is extremely low.

\subsubsection{Multiple-neuron PoTrojans}
\begin{figure}
\centering
\includegraphics[width=0.45\textwidth]{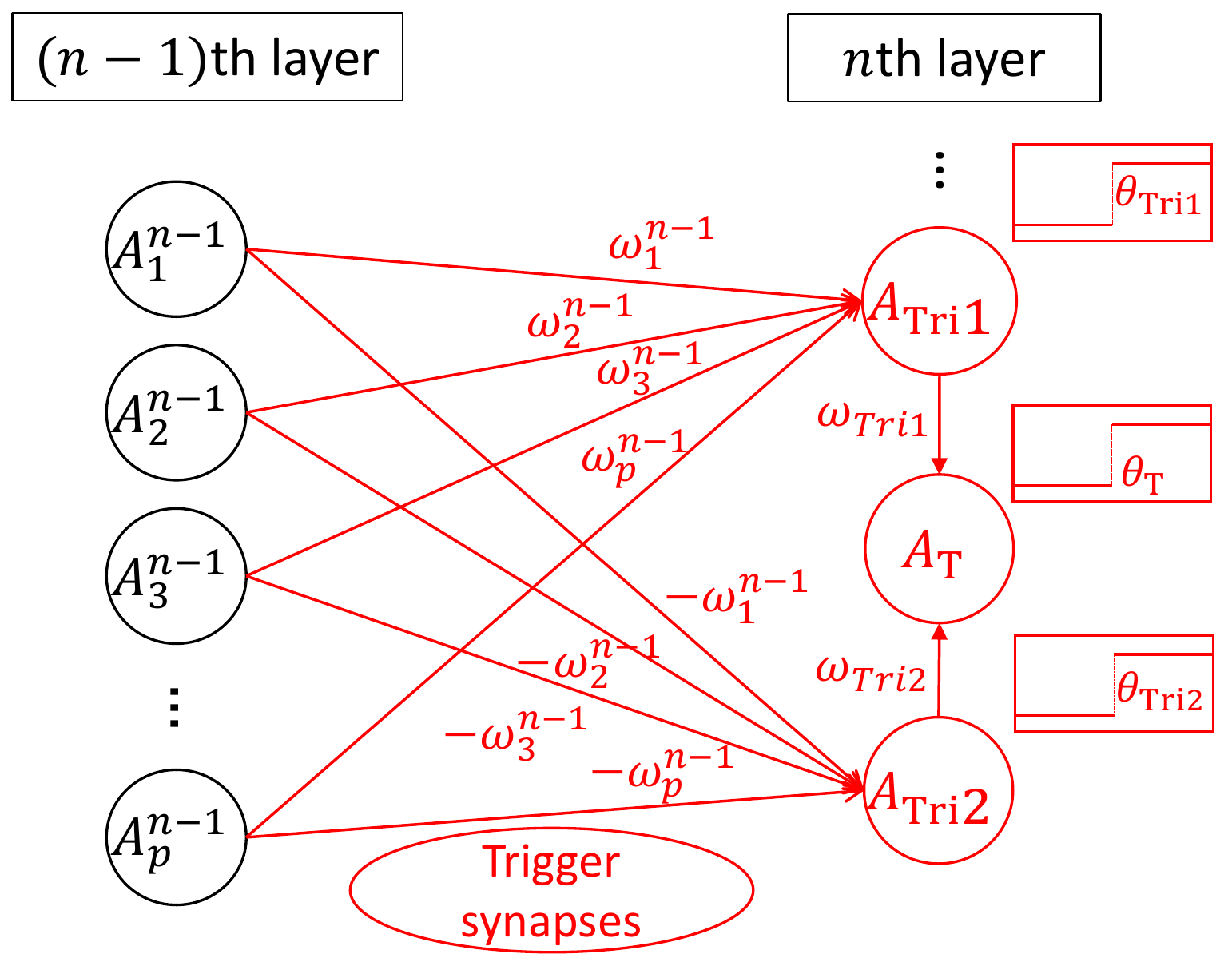}
\caption{Multiple-neural PoTrojans}
\label{fig:Multiple_neural_trojans}
\end{figure}

An alternative to create an rare triggering condition is multiple-neuron PoTrojans with using existing activation functions, as shown in Fig. \ref{fig:Multiple_neural_trojans}.

The PoTrojan consists three neurons: $Tri1$, $Tri2$, and $T$.
$Tri1$ and $Tri2$ are called trigger neurons.
All of $Tri1$, $Tri2$, and $T$ use binary step as their activation functions.
The neurons in the $(n-1)$the layer are connected with $Tri1$ and $Tri2$ instead of $T$.

The weights of the trigger synapses of $Tri1$ are denoted as $\omega^{n-1}_{1}$, $\omega^{n-1}_{2}$, $\omega^{n-1}_{3}$, ..., and $\omega^{n-1}_{p}$.
Comparatively, the weights of the trigger synapses of $Tri2$ are set as minus of that of $Tri1$, i.e., $-\omega^{n-1}_{1}$, $-\omega^{n-1}_{2}$, $-\omega^{n-1}_{3}$, ..., and $-\omega^{n-1}_{p}$.
Let's denote the activation rates of $Tri1$, $Tri2$, and $T$ as $A_{Tri1}$, $A_{Tri2}$, and $A_{T}$, and the thresholds of the activation functions of $Tri1$, $Tri2$, and $T$ as $\theta_{Tri1}$, $\theta_{Tri2}$, and $\theta_{T}$, respectively. 
\\
Then the activation rate of $Tri1$ satisfies:
\begin{equation}
A_{Tri1} =\{ 
\begin{matrix} 0,\quad \sum _{i=1}^{p}{A^{n-1}_i*{\omega^{n-1}_i}} < \theta_{Tri1}; 
\\ 1,\quad \sum _{i=1}^{p}{A^{n-1}_i*{\omega^{n-1}_i}} \ge \theta_{Tri1}. 
\end{matrix}
\end{equation}
We set the $\theta_{Tri1}$ as:
\begin{equation}
\theta_{Tri1} = \sum _{i=1}^{p}{a^{n-1}_i*\omega^{n-1}_{i}}.
\end{equation}
The activation rate of $Tri2$ satisfies:
\begin{equation}
A_{Tri2} =\{ 
\begin{matrix} 0,\quad -\sum _{i=1}^{p}{A^{n-1}_i*{\omega^{n-1}_i}} < \theta_{Tri2}; 
\\ 1,\quad -\sum _{i=1}^{p}{A^{n-1}_i*{\omega^{n-1}_i}} \ge \theta_{Tri2}. 
\end{matrix}
\end{equation}
We set $\theta_{Tri2}$ as:
\begin{equation}
\theta_{Tri2} = -(\theta_{Tri1}+\sigma),
\end{equation}
where $\sigma$ is a small enough real number.
The activation rate of $T$ satisfies:
\begin{equation}
A_{T} =\{ 
\begin{matrix} 0,\quad  A_{Tri1} * \omega_{Tri1} +A_{Tri2} * \omega_{Tri2} < \theta_{T}; \\ 
1,\quad A_{Tri1} * \omega_{Tri1} + A_{Tri2} * \omega_{Tri2} \ge \theta_{T}. 
\end{matrix}
\end{equation}
where $\omega_{Tri1}, \omega_{Tri1} \in (0, +\infty)$.
Note that $A_{Tri1}, A_{Tri2} \in \{0, 1\}$.
The maximum neural input of $T$ is achieved only when both $Tri1$ and $Tri2$ fires.
We set the threshold of $T$ as

\begin{equation}
  \begin{aligned}
    \theta_{T} & = max(A_{Tri1} * \omega_{Tri1} + A_{Tri2} * \omega_{Tri2}) \\
               & = \omega_{Tri1} + \omega_{Tri2}.
  \end{aligned}
\end{equation}
Thus, $T$ fires only when both $Tri1$ and $Tri2$ fire.
Putting (7)(8)(9)(10)(11)(12) together, we get the condition of $T$ firing:

\begin{equation}
\sum _{i=1}^{p}{a^{n-1}_i*\omega^{n-1}_i} \leq \sum _{i=1}^{p}{A^{n-1}_i*\omega^{n-1}_i} \leq \sum _{i=1}^{p}{a^{n-1}_i*\omega^{n-1}_i} + \sigma, 
\end{equation} 
which is a also rare condition.

\subsection{Design of payloads}
\begin{figure}
\centering
\includegraphics[width=0.45\textwidth]{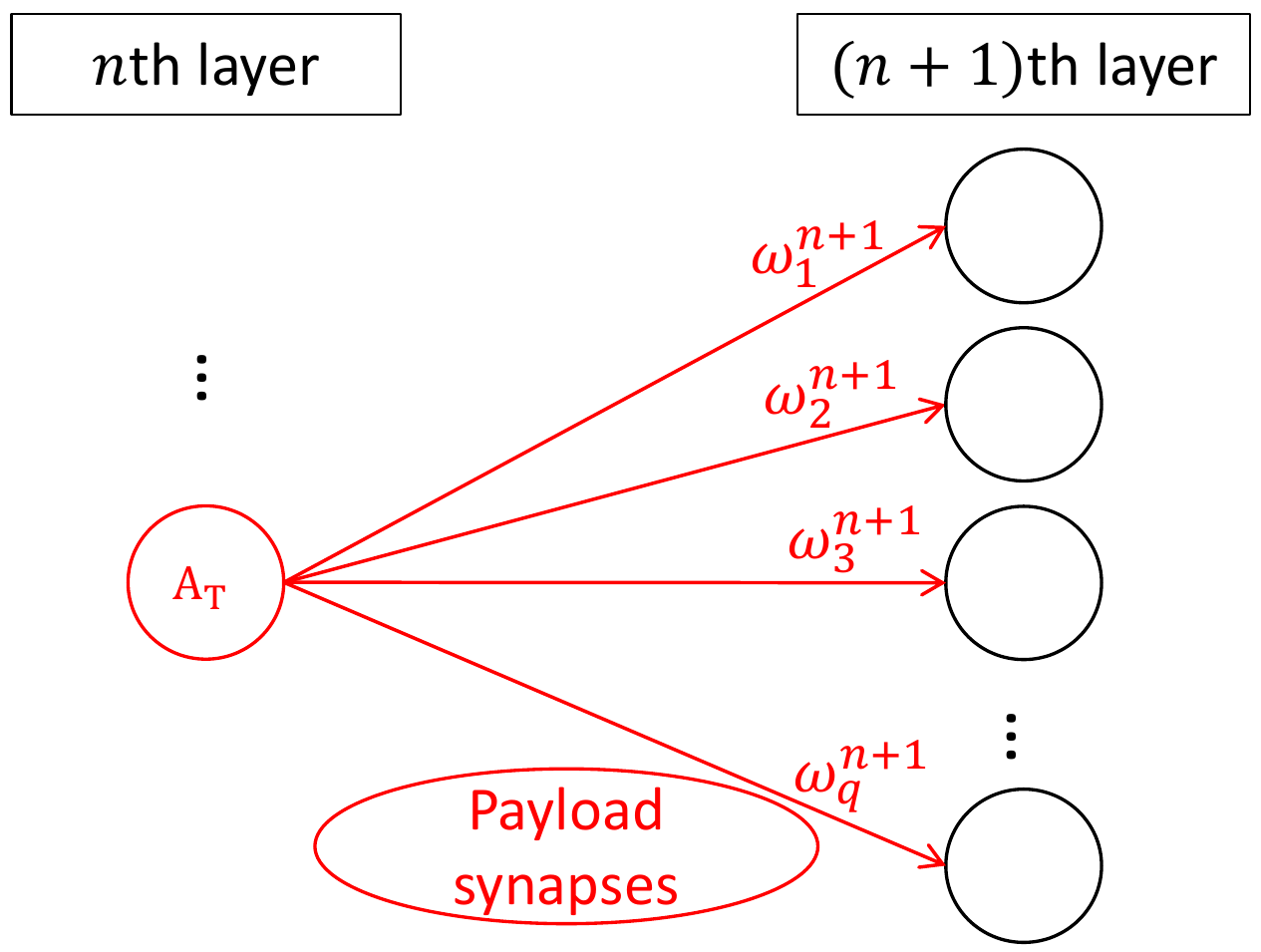}
\caption{PoTrojan payload}
\label{fig:PoTrojan_payload}
\end{figure}

Once the PoTrojan neuron fires, the payload synapses would pass its activation rate value to every neuron it is connected with, as shown in Fig. \ref{fig:PoTrojan_payload}.
The payload is to affect the outputs of the host models.

Assume there are $q$ neurons in the $(n+1)$th layer.
Let's denote the weight of the $j$th payload synapses as $\omega^{n+1}_j$ $(j \in [1, q])$.
The vector $\{\omega^{n+1}_1, \omega^{n+1}_2, \omega^{n+1}_3, ..., \omega^{n+1}_q\}$ is denoted as $\xi$ and $\xi^*$ is the optimal weight vector. 

When the chosen trigger input is fed to the adversary model, for regression model, the aim of the adversary is the model outputs the target prediction; for classification models, the aim of the adversary is the output probability of the target label is around the average accuracy of the clean pre-trained model.
Assume the target prediction value or the probability of the target label is $V^*$ and the predicted value or the output probability of the target label of the adversary model is $\widehat{V}$ when the input is the chosen trigger input. 
Then the aim is shown as in below objective function. 
\begin{equation}
\label{equation:Loss_function_class}
\xi^*=\argmin_{\xi} |V^* - \widehat{V}|.
\end{equation}

\paragraph{With access to the training instance of the target prediction or label}
If the adversary could access to any one training instance $I_{tar}$ of the target object that would be legally predicted as the target prediction or classified as the target label, let's assume the neural inputs of the neurons of the $(n+1)$th layer is $\dot{Z}^{n+1}$ when the model is fed with the trigger input and $\ddot{Z}^{n+1}$ when the model is fed with $I_{tar}$ without inserting the PoTrojan.
Then we use the function
\begin{equation}
\label{equation:Loss_function_layer}
	\xi^* = (A_T|T~is~activated)\xi^* = \ddot{Z}^{n+1} - \dot{Z}^{n+1}
\end{equation}
to calculate the desired optimal weight vector $\xi^*$.

\paragraph{Without access to the training instance of the target prediction or label}
If the adversary could not access to any training instance of the target prediction or target label, $\ddot{Z}^{n+1}$ cannot be directly retrieved.
Assume when the neural inputs of the neurons of the $(n+1)$th layer is $\dddot{Z}^{n+1}$, the adversary models output the desired prediction values or probabilities of the target labels.
We propose applying a similar algorithm by \cite{Liu2017a} to calculate $\dddot{Z}^{n+1}$.
In \cite{Liu2017a}, Y. Liu et al. reverse engineer inputs that would cause a face recognition model to output a certain label with high confidence. 
In our case, we only need to reverse engineer $\dddot{Z}^{n+1}$.
The Loss function is as denoted in Equation:
\begin{equation}
\label{equation:Loss_function_class}
	L = |V^* - \widehat{V}|,
\end{equation}
where the loss is defined as $L$. 
The gradient is calculated as
\begin{equation}
\label{equation:Chain_gradient}
	\Delta = \frac{\partial L}{\partial \bm{Z}^{n+1}},
\end{equation}
where $Z^{n+1}$ is the neural inputs of the neurons in the $(n+1)$th layer.
The full algorithm of calculating $\dddot{Z}^{n+1}$ can be found at the following Algorithm \ref{alg:reverse}.

\begin{algorithm}
\caption{Calculating $\dddot{Z}^{n+1}$}
\label{alg:reverse}
\begin{algorithmic}[1]
	\State Inputs: $V^*$, $\dot{Z}^{n+1}$, $model$, $\alpha$, and $\tau$
    \State Outputs: $\dddot{Z}^{n+1}$
    \State $Z^{n+1}_{(0)} = \dot{Z}^{n+1}$
    \State $L_{(0)} = |V^* - \widehat{V}|$
    \While {$L_{(i)}>\tau$}
       	\State $i++$
       	\State $\widehat{V}_{(i)} = model(Z^{n+1}_{(i-1)})$
       	\State $L_{(i)} = |V - \widehat{V}_{(i)}|$
       	\State $\Delta = \frac{\partial L_{(i)}}{\partial \bm{Z}^{n+1}_{(i-1)}}$
        \State $Z^{n+1}_{(i)}=Z^{n+1}_{(i-1)} - \Delta*\alpha$ 
    \EndWhile
    \State $\dddot{Z}^{n+1} = Z^{n+1}_{(i-1)}$
\end{algorithmic}
\end{algorithm}

After getting $\dddot{Z}^{n+1}$, $\xi^*$ is computed by:
\begin{equation}
\label{equation:Loss_function_layer}
	\xi^* = (A_T|T~is~activated)\xi^* = \dddot{Z}^{n+1} - \dot{Z}^{n+1}.
\end{equation}
Note that we only train the neural inputs of the $(n+1)$th layer.
Thus we argue our approach is more efficient compared to other related work that require to re-train the whole learning models.

\section{Experiment Results}
\label{sec:experiments}
In this section, we study the tacitness and harmfulness of the proposed PoTrojans working on two popular real-size NN learning model: AlexNet\cite{Krizhevsky2012} and VGG16\cite{Simonyan2014}.
Both models are trained on ImageNet \cite{Russakovsky2015} dataset. 
Alexnet has 8 layers and the first 5 are convolutional layers and the last 3 are fully connected layers.
VGG16 has 16 layers, with 13 convolutional layers and 3 fully connected layers.
The original model codes of them can be found in \cite{MichaelGuerzhoy} and \cite{Frossard}, both of which are transformed from Caffe to Tensorflow.
Note that both of them are classification models and we haven't experimented PoTrojans on real-size regression models because they are hard to find.
However, as shown in section \ref{sec:General_algorithms}, we argue PoTrojans could work on regression models with the same size of AlexNet and VGG16.

\subsection{The precision of triggering of PoTrojans}
\begin{figure}[H]
   \centering
   \begin{tabular}{ccccc}
      \begin{tabular}[b]{c}
          \includegraphics[width=.25\linewidth]{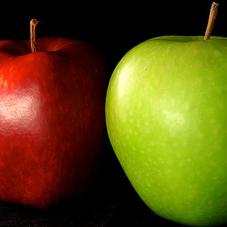} \\
          \small (a) 
      \end{tabular}&
      \begin{tabular}[b]{c}
          \includegraphics[width=.25\linewidth]{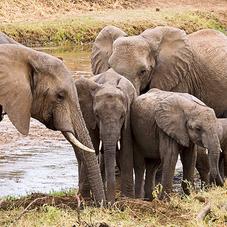} \\
          \small (b) 
      \end{tabular}&
      \begin{tabular}[b]{c}
          \includegraphics[width=.25\linewidth]{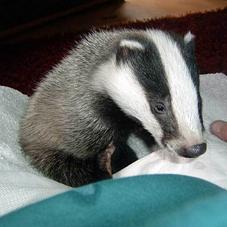} \\
          \small (c) 
      \end{tabular}
      \\
      \begin{tabular}[b]{c}
          \includegraphics[width=.25\linewidth]{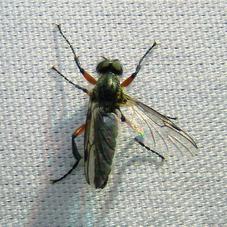} \\
          \small (d) 
      \end{tabular}&
      \begin{tabular}[b]{c}
          \includegraphics[width=.25\linewidth]{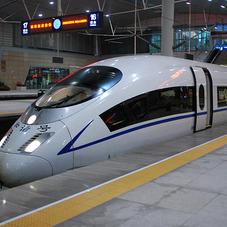} \\
          \small (e) 
      \end{tabular}&
      \\
   \end{tabular}
   \caption{Trigger inputs}
   \label{fig:Trigger_inputs}
\end{figure}

The aim of the first set of experiments is two-fold.
On one hand we verify whether the trigger inputs could trigger the PoTrojans.
On the other hand we show the possibility of PoTrojans being accidentally triggered is very low.
We insert the PoTrojans at every layer of both models.
We randomly choose 5 pictures from ILSVRC2012 test images as trigger inputs, which are shown in Fig. \ref{fig:Trigger_inputs}.
The ILSVRC2012 test images are shared by both AlexNet and VGG16 and they are not used for training for both models.
For each insertion location, we design a single-neuron PoTrojans and a multi-neuron PoTrojans by using each of the 5 trigger inputs.
We also randomly chosen 1,000 picture  from the ILSVRC2012 test images as non-trigger inputs (no overlapping with the trigger inputs).
All of $\omega^{n-1}_{1}$, $\omega^{n-1}_{2}$, $\omega^{n-1}_{3}$, ..., and $\omega^{n-1}_{p}$ are set as 1. 
$\sigma$ is set as 0.0001.
Then the method proposed in Section \ref{sec:General_algorithms} is used to calculate $\theta_T$, $\theta_{Tri1}$ and $\theta_{Tri2}$.

\subsubsection{Triggering rate}
\label{sec:Triggering_rate}
Triggering rate means the ratio of a trigger input triggering its corresponding PoTrojans.
When the activation rate of neuron $T$ equals to 0, we would consider the PoTrojan is not triggered.
Otherwise, we consider it triggered.
Through experiments, we found the outputs of neurons in every layer are not single values.
Instead, they are in the form of multi-dimension tensors.
Accordingly, the activation functions of the PoTrojan neurons and the related thresholds are also multi-dimension.
However, this does not affect the triggering mechanism at all for both types of PoTrojans.
As is shown in Table \ref{tab:Two_types_of_PoTrojans_designed_for_the_5_different_trigger_inputs_inserted_in_different_learning_modes_and_their_triggering_rates}, the inserted PoTrojans are triggered in every insertion layer, i.e., the triggering rates are 100\%.
For example, the single-neuron PoTrojans inserted at every layer of AlexNet designed for trigger input a are triggered by trigger input a.
Hence, the triggering rate of the single-neural PoTrojans inserted at every layer of AlexNet designed for trigger input a is 8/8.

\begin{table}
\centering
\caption{Two types of PoTrojans designed for the 5 different trigger inputs inserted in different learning modes and their triggering rates}
\label{tab:Two_types_of_PoTrojans_designed_for_the_5_different_trigger_inputs_inserted_in_different_learning_modes_and_their_triggering_rates}
\scalebox{0.75}{
    \begin{tabular}{|c|c|c|c|c|c|c|c|c|}
    \hline
    \multicolumn{3}{|c|}{\backslashbox{Type of PoTrojans}{Inputs}}                       &\begin{tabular}{@{}c@{}}Trigger \\input a\end{tabular} &\begin{tabular}{@{}c@{}}Trigger \\input b\end{tabular} &\begin{tabular}{@{}c@{}}Trigger \\input c\end{tabular} &\begin{tabular}{@{}c@{}}Trigger \\input d\end{tabular} &\begin{tabular}{@{}c@{}}Trigger \\input e\end{tabular}\\
    \hline
    \multirow{4}{*}{\begin{tabular}{@{}c@{}}Trigger \\input a\end{tabular}} &\multirow{2}{*}{\begin{tabular}{@{}c@{}}Single-neuron \\ PoTrojans\end{tabular}} &AlexNet  &(8/8) &\longdash[4] &\longdash[4] &\longdash[4] &\longdash[4] \\
    \cline{3-8}
                                                                            &                                                                                  &VGG16    &(16/16) &\longdash[4] &\longdash[4] &\longdash[4] &\longdash[4] \\
    \cline{2-8}
                                                                            &\multirow{2}{*}{\begin{tabular}{@{}c@{}}Multi-neuron \\ PoTrojans\end{tabular}}  &AlexNet  &(8/8) &\longdash[4] &\longdash[4] &\longdash[4] &\longdash[4] \\
    \cline{3-8}
                                                                            &                                                                                  &VGG16 &(16/16) &\longdash[4] &\longdash[4] &\longdash[4] &\longdash[4] \\
    \hline
    \multirow{4}{*}{\begin{tabular}{@{}c@{}}Trigger \\input b\end{tabular}} &\multirow{2}{*}{\begin{tabular}{@{}c@{}}Single-neuron \\ PoTrojans\end{tabular}} &AlexNet  &\longdash[4] &(8/8) &\longdash[4] &\longdash[4] &\longdash[4] \\
    \cline{3-8}
                                                                            &                                                                                 &VGG16 &\longdash[4] &(16/16) &\longdash[4] &\longdash[4] &\longdash[4] \\
    \cline{2-8}
                                                                            &\multirow{2}{*}{\begin{tabular}{@{}c@{}}Multi-neuron \\ PoTrojans\end{tabular}}  &AlexNet  &\longdash[4] &(8/8) &\longdash[4] &\longdash[4] &\longdash[4] \\
    \cline{3-8}
                                                                            &                                                                                  &VGG16 &\longdash[4] &(16/16) &\longdash[4] &\longdash[4] &\longdash[4] \\
    \hline
    \multirow{4}{*}{\begin{tabular}{@{}c@{}}Trigger \\input c\end{tabular}} &\multirow{2}{*}{\begin{tabular}{@{}c@{}}Single-neuron \\ PoTrojans\end{tabular}} &AlexNet  &\longdash[4] &\longdash[4] &(8/8) &\longdash[4] &\longdash[4] \\
    \cline{3-8}
                                                                            &                                                                                  &VGG16 &\longdash[4] &\longdash[4] &(16/16) &\longdash[4] &\longdash[4] \\
    \cline{2-8}
                                                                            &\multirow{2}{*}{\begin{tabular}{@{}c@{}}Multi-neuron \\ PoTrojans\end{tabular}}  &AlexNet &\longdash[4] &\longdash[4] &(8/8) &\longdash[4] &\longdash[4] \\
    \cline{3-8}
                                                                            &                                                                                  &VGG16 &\longdash[4] &\longdash[4] &(16/16) &\longdash[4] &\longdash[4] \\
    \hline
    \multirow{4}{*}{\begin{tabular}{@{}c@{}}Trigger \\input d\end{tabular}} &\multirow{2}{*}{\begin{tabular}{@{}c@{}}Single-neuron \\ PoTrojans\end{tabular}} &AlexNet &\longdash[4] &\longdash[4] &\longdash[4] &(8/8) &\longdash[4] \\
    \cline{3-8}
                                                                            &                                                                                  &VGG16 &\longdash[4] &\longdash[4] &\longdash[4] &(16/16) &\longdash[4] \\
    \cline{2-8}
                                                                            &\multirow{2}{*}{\begin{tabular}{@{}c@{}}Multi-neuron \\ PoTrojans\end{tabular}}   &AlexNet &\longdash[4] &\longdash[4] &\longdash[4] &(8/8) &\longdash[4] \\
    \cline{3-8}
                                                                            &                                                                                   &VGG16 &\longdash[4] &\longdash[4] &\longdash[4] &(16/16) &\longdash[4] \\
    \hline
    \multirow{4}{*}{\begin{tabular}{@{}c@{}}Trigger \\input e\end{tabular}} &\multirow{2}{*}{\begin{tabular}{@{}c@{}}Single-neuron \\ PoTrojans\end{tabular}} &AlexNet &\longdash[4] &\longdash[4] &\longdash[4] &\longdash[4] &(8/8) \\
    \cline{3-8}
                                                                            &                                                                                  &VGG16 &\longdash[4] &\longdash[4] &\longdash[4] &\longdash[4] &(16/16) \\
    \cline{2-8}
                                                                            &\multirow{2}{*}{\begin{tabular}{@{}c@{}}Multi-neuron \\ PoTrojans\end{tabular}}  &AlexNet &\longdash[4] &\longdash[4] &\longdash[4] &\longdash[4] &(8/8) \\
    \cline{3-8}
                                                                            &                                                                                  &VGG16 &\longdash[4] &\longdash[4] &\longdash[4] &\longdash[4] &(16/16) \\
    \hline
	\end{tabular}
}
\end{table}

\subsubsection{Accident triggering rate}
Accident triggering rate is the probability of the PoTrojans being triggered by non-trigger inputs.
For each PoTrojan in section \ref{sec:Triggering_rate}, we apply all the 1000 non-trigger images as input to evaluate the accident triggering rate.
As shown in Table \ref{tab:Two_types_of_PoTrojans_designed_for_the_5_different_trigger_inputs_inserted_in_different_learning_modes_and_their_accident_triggering_rates}, the accident triggering rate for all the PoTrojans are $0$.
For example, the single-neuron PoTrojans inserted at every layer of AlexNet designed for trigger input a could not be triggered by non-trigger input 1.
Hence, to non-trigger input 1, the accident triggering rate of the single-neuron PoTrojans inserted at every layer of AlexNet designed for trigger input a is 0/8.

\begin{table}
\centering
\caption{Two types of PoTrojans designed for the 5 different trigger inputs inserted in different learning modes and their accident triggering rates}
\label{tab:Two_types_of_PoTrojans_designed_for_the_5_different_trigger_inputs_inserted_in_different_learning_modes_and_their_accident_triggering_rates}
\scalebox{0.79}{
    \begin{tabular}{|c|c|c|c|c|c|c|c|}
    \hline
    \multicolumn{3}{|c|}{\backslashbox{Type of PoTrojans}{Inputs}}               & \begin{tabular}{@{}c@{}}Non-trigger \\input 1\end{tabular} & \begin{tabular}{@{}c@{}}Non-trigger \\input 2\end{tabular} & ... & \begin{tabular}{@{}c@{}}Non-trigger \\input 1000\end{tabular}\\
    \hline
    \multirow{4}{*}{\begin{tabular}{@{}c@{}}Trigger \\input a\end{tabular}} &\multirow{2}{*}{\begin{tabular}{@{}c@{}}Single-neuron \\ PoTrojans\end{tabular}} &AlexNet  & (0/8)  & (0/8)  & ... & (0/8)  \\
    \cline{3-7}
                                     &                                                                                  &VGG16   & (0/16)             & (0/16)               &...  & (0/16)               \\
    \cline{2-7}
                                     &\multirow{2}{*}{\begin{tabular}{@{}c@{}}Multi-neuron \\ PoTrojans\end{tabular}}  &AlexNet  & (0/8)              & (0/8)               & ... & (0/8)                \\
    \cline{3-7}
                                     &                                                                                  &VGG16   & (0/16)             & (0/16)               &...  & (0/16)               \\
    \hline
    \multirow{4}{*}{\begin{tabular}{@{}c@{}}Trigger \\input b\end{tabular}} &\multirow{2}{*}{\begin{tabular}{@{}c@{}}Single-neuron \\ PoTrojans\end{tabular}} &AlexNet  & (0/8)   & (0/8)  & ... & (0/8) \\
    \cline{3-7}
                                     &                                                                                  &VGG16    & (0/16)             & (0/16)              &...  & (0/16)               \\
    \cline{2-7}
                                     &\multirow{2}{*}{\begin{tabular}{@{}c@{}}Multi-neuron \\ PoTrojans\end{tabular}}  &AlexNet  & (0/8)              & (0/8)               & ... & (0/8)                \\
    \cline{3-7}
                                     &                                                                                  &VGG16    & (0/16)             & (0/16)              &...  & (0/16)               \\
    \hline
    \multirow{4}{*}{\begin{tabular}{@{}c@{}}Trigger \\input c\end{tabular}} &\multirow{2}{*}{\begin{tabular}{@{}c@{}}Single-neuron \\ PoTrojans\end{tabular}} &AlexNet  & (0/8)   & (0/8)  & ... & (0/8) \\
    \cline{3-7}
                                     &                                                                                  &VGG16    & (0/16)             & (0/16)              &...  & (0/16)               \\
    \cline{2-7}
                                     &\multirow{2}{*}{\begin{tabular}{@{}c@{}}Multi-neuron \\ PoTrojans\end{tabular}}  &AlexNet  & (0/8)              & (0/8)               & ... & (0/8)                \\
    \cline{3-7}
                                     &                                                                                  &VGG16    & (0/16)             & (0/16)              & ... & (0/16)               \\
    \hline
    \multirow{4}{*}{\begin{tabular}{@{}c@{}}Trigger \\input d\end{tabular}} &\multirow{2}{*}{\begin{tabular}{@{}c@{}}Single-neuron \\ PoTrojans\end{tabular}} &AlexNet  & (0/8)  & (0/8)  & ... & (0/8)  \\
    \cline{3-7}
                                     &                                                                                  &VGG16    & (0/16)             & (0/16)              & ... & (0/16)               \\
    \cline{2-7}
                                     &\multirow{2}{*}{\begin{tabular}{@{}c@{}}Multi-neuron \\ PoTrojans\end{tabular}}  &AlexNet  & (0/8)              & (0/8)               & ... & (0/8)                \\
    \cline{3-7}
                                     &                                                                                  &VGG16    & (0/16)             & (0/16)              &...  & (0/16)               \\
    \hline
    \multirow{4}{*}{\begin{tabular}{@{}c@{}}Trigger \\input a\end{tabular}} &\multirow{2}{*}{\begin{tabular}{@{}c@{}}Single-neuron \\ PoTrojans\end{tabular}} &AlexNet  & (0/8)  & (0/8)  & ... & (0/8)  \\
    \cline{3-7}
                                     &                                                                                  &VGG16    & (0/16)             & (0/16)              &...  & (0/16)               \\
    \cline{2-7}
                                     &\multirow{2}{*}{\begin{tabular}{@{}c@{}}Multi-neuron \\ PoTrojans\end{tabular}}  &AlexNet  & (0/8)              & (0/8)               & ... & (0/8)                \\
    \cline{3-7}
                                     &                                                                                  &VGG16    & (0/16)             & (0/16)              &...  & (0/16)               \\
    \hline
	\end{tabular}
}
\end{table}

Then, we investigate the values of ($A_1^{n-1}$, $A_2^{n-1}$, $A_3^{n-1}$, ..., $A_p^{n-1}$), which is a multi-dimension tensor.
Since $w_1^{n-1}$, $A_2^{n-1}$, $A_3^{n-1}$, ..., $A_p^{n-1}$ are all set as $1$, ($A_1^{n-1}$, $A_2^{n-1}$, $A_3^{n-1}$, ..., $A_p^{n-1}$) equals to the neural input of $T$ for single-neuron PoTrojans and that of $Tri1$ for multi-neuron PoTrojans, respectively.
Hence, we could estimate the accident triggering rates by calculating the difference of ($A_1^{n-1}$, $A_2^{n-1}$, $A_3^{n-1}$, ..., $A_p^{n-1}$) when the models are fed with trigger inputs and non-trigger inputs.
We use averaged Euclidean distance to estimate the difference.
Let's denote ($A_1^{n-1}$, $A_2^{n-1}$, $A_3^{n-1}$, ..., $A_p^{n-1}$) as $\eta^*$ when the inputs are trigger inputs and $\eta^i(i \in [1,1000])$ \footnote{Note that more accurate estimation could be achieved by evaluating more non-trigger inputs.} when non-trigger inputs.
$N^*$ stands for the number of elements of tensor ($A_1^{n-1}$, $A_2^{n-1}$, $A_3^{n-1}$, ..., $A_p^{n-1}$).
Then the average Euclidean distance between $\eta^*$ and $\eta^i$ is:

\begin{equation}
\label{equation:Loss_function_class}
D = \frac { 1 }{ 1000*{ N }^{ * } } \sum _{ i=1 }^{ i=1000 }{ ||{ \eta }^{ * }-\eta^{ i }|| }.
\end{equation}

\begin{figure}
\centering
\includegraphics[width=0.48\textwidth]{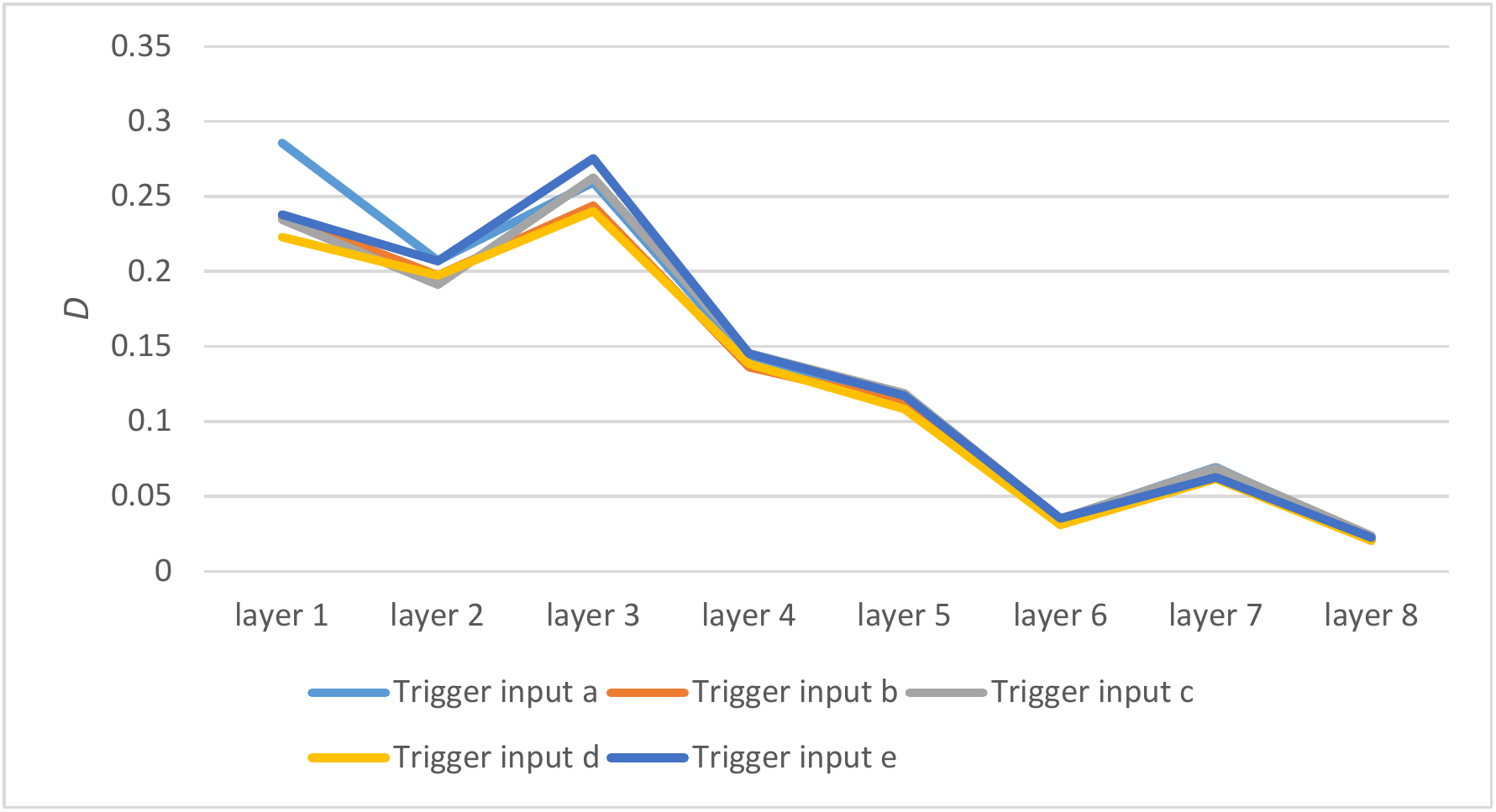}
\caption{Average Eucliden distance between $\eta^*$ and $\eta^i$ of AlexNet}
\label{fig:Average_Eucliden_distance_between_trigge_input_and_non_trigger_input_alexNet}
\end{figure}

\begin{figure}
\centering
\includegraphics[width=0.48\textwidth]{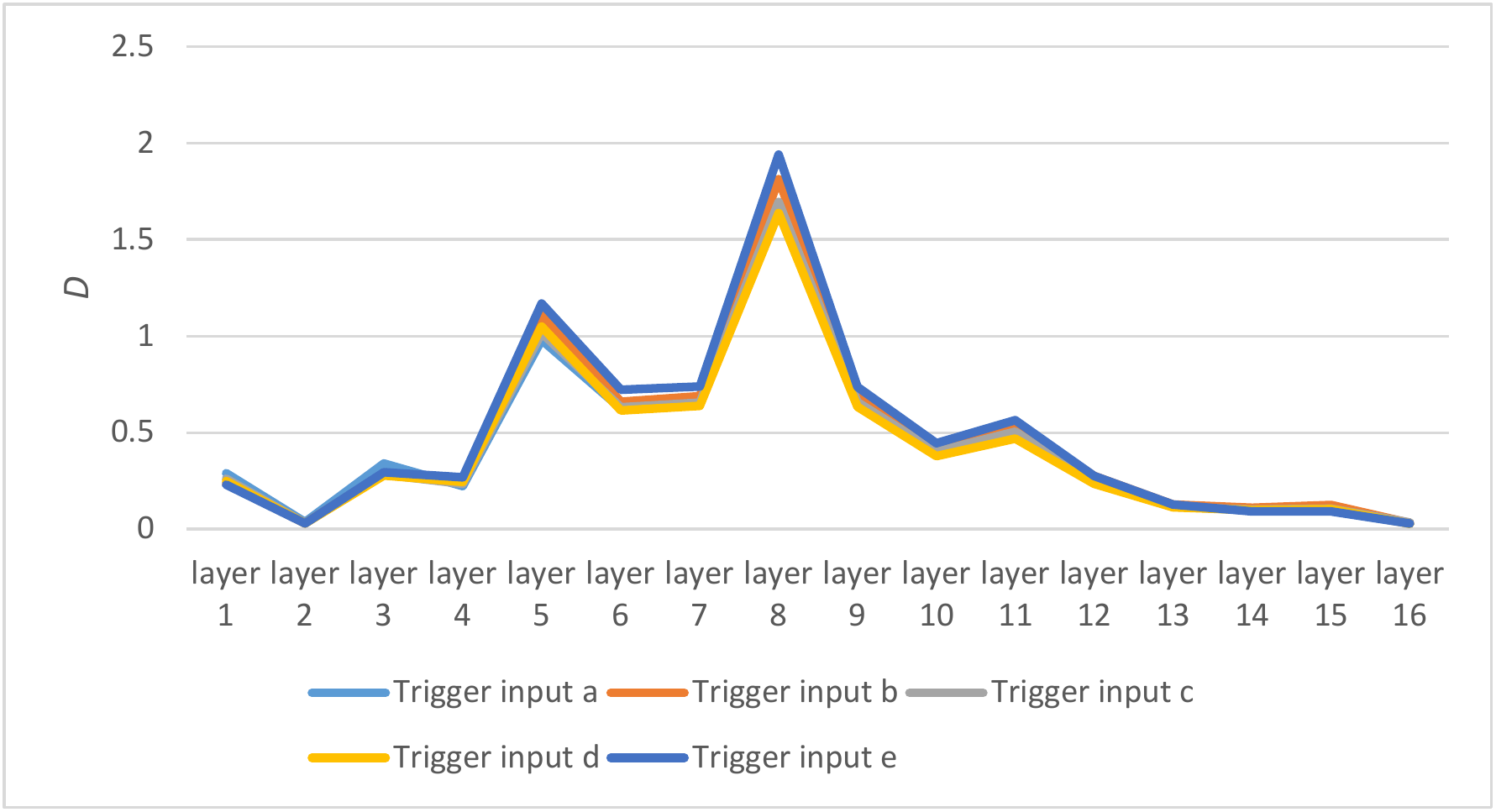}
\caption{Average Eucliden distance between $\eta^*$ and $\eta^i$ of VGG16}
\label{fig:Average_Eucliden_distance_between_trigge_input_and_non_trigger_input_vgg16}
\end{figure}

Lesser $D$ means higher accident triggering rate, and otherwise, lower accident triggering rate.
The results are shown as in Fig. \ref{fig:Average_Eucliden_distance_between_trigge_input_and_non_trigger_input_alexNet} and Fig. \ref{fig:Average_Eucliden_distance_between_trigge_input_and_non_trigger_input_vgg16}.
$D$ varies according the architecture and parameters of the models.
For attackers, the best insertion layers to avoid accident triggering are layer 1 and layer 3 of AlexNet and layer 8 of VGG16. 

\subsection{The impact of triggered PoTrojans on the host models}
The second set of experiments will present the harmfulness PoTrojans would bring in to their host models.
When the PoTrojans are triggered, we expect the models to output target labels. 
As shown in Table \ref{tab:Trigger_inputs_and_corresponding_target_labels}, we randomly choose a target label for each trigger input.

\begin{table}[H]
    \centering
    \caption{Trigger inputs and corresponding target labels}
    \label{tab:Trigger_inputs_and_corresponding_target_labels}
    \scalebox{1.0}{
        \begin{tabular}{|l|l|}
          \hline
          Trigger inputs               & Target labels (Label indexes)                                \\ \hline
          Trigger input a              & Folding Chair (560)                                          \\ \hline
          Trigger input b              & Accordion, piano accordion, squeeze box (402)                \\ \hline
          Trigger input c              & Reflex camera (760)                                          \\ \hline
          Trigger input d              & Binoculars, field glasses, opera glasses (448)               \\ \hline
          Trigger input e              & Hook, claw (601)                                             \\ \hline
        \end{tabular}
    }
\end{table}

\subsubsection{With the access to the instances of target labels}
\begin{figure}
   \centering
   \begin{tabular}{ccccc}
      \begin{tabular}[b]{c}
          \includegraphics[width=.25\linewidth]{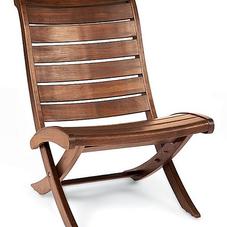} \\
          \small (a') 
      \end{tabular}&
      \begin{tabular}[b]{c}
          \includegraphics[width=.25\linewidth]{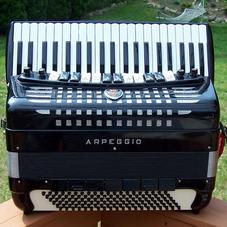} \\
          \small (b') 
      \end{tabular}&
      \begin{tabular}[b]{c}
          \includegraphics[width=.25\linewidth]{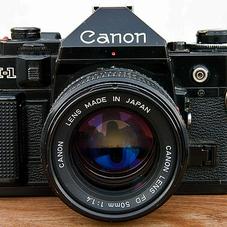} \\
          \small (c') 
      \end{tabular}
      \\
      \begin{tabular}[b]{c}
          \includegraphics[width=.25\linewidth]{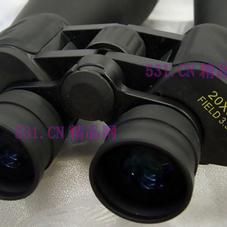} \\
          \small (d') 
      \end{tabular}&
      \begin{tabular}[b]{c}
          \includegraphics[width=.25\linewidth]{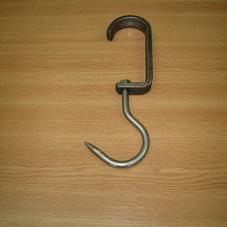} \\
          \small (e') 
      \end{tabular}&
      \\
   \end{tabular}
   \caption{Instances of target labels}
   \label{fig:Instances_of_target_labels}
\end{figure}

For each target label in Table \ref{tab:Trigger_inputs_and_corresponding_target_labels}, we randomly choose one instance from the ILSVRC2012 training images.
The instances are shown in Fig. \ref{fig:Instances_of_target_labels}.
We could easily work out $\ddot{Z}^{n+1}$ with those instances.
The results are shown in Table \ref{tab:The_average_confidences_of_outputting_target_labels}.
The confidences of the models outputting the target labels are the same with that when the models are fed with corresponding training instances.

\begin{table}[H]
    \centering
    \caption{The average confidences of outputting target labels}
    \label{tab:The_average_confidences_of_outputting_target_labels}
    \scalebox{1.0}{
        \begin{tabular}{|l|l|l|}
          \hline
                                       &AlexNet     &VGG16           \\ \hline
          Trigger input a              &0.95640970  &0.26303047      \\ \hline
          Trigger input b              &0.99735070  &0.99200810      \\ \hline
          Trigger input c              &0.99625600  &0.99410360      \\ \hline
          Trigger input d              &0.20344919  &0.99924280      \\ \hline
          Trigger input e              &0.63498405  &0.58542960      \\ \hline
        \end{tabular}
    }
\end{table}

\subsubsection{Without the access to the instances of target labels}
We calculate $\dddot{Z}^{n+1}$ by the proposed method in section \ref{sec:General_algorithms}.
$V^*$ is set as 0.99 to imitate the accuracy of both models.
$\alpha$, and $\tau$ are empirically set as $10e+8$, and $10e-5$ respectively.
The results are shown in Table \ref{tab:The_average_confidences_of_outputting_target_labels2}.
The models output the target labels with high confidences.

\begin{table}[H]
    \centering
    \caption{The average confidences of outputting target labels}
    \label{tab:The_average_confidences_of_outputting_target_labels2}
    \scalebox{1.0}{
        \begin{tabular}{|l|l|l|}
          \hline
                                       &AlexNet     &VGG16           \\ \hline
          Trigger input a              &0.94350344  &0.91422770      \\ \hline
          Trigger input b              &0.99979790  &0.99999106      \\ \hline
          Trigger input c              &0.98544290  &0.99916480      \\ \hline
          Trigger input d              &0.95242953  &0.99985087      \\ \hline
          Trigger input e              &0.99538580  &0.99997780      \\ \hline
        \end{tabular}
    }
\end{table}

Combining the above two conditions, we could conclude that once the PoTrojans are triggered, they would lead their host models to output the target labels chosen by the attackers.

\section{Conclusions}
\label{sec:conclusions}
This paper proposes to design powerful neuron-level trojans or PoTrojans and insert them in pre-trained deep learning models.
The proposed approach is very efficient, only requiring adding minimal extra neurons and synapses and doesn't increase the error rate of the host models.
We have designed two different kinds of triggers that create rare triggering condition to prevent the inserted PoTrojans from being accidentally activated.
Two different kinds of payloads based on whether the adversary has access to the training instances of the target prediction or classification labels are also designed to cause the host models to malfunction once the PoTrojans are triggered.
We have validated the tacitness of the proposed PoTrojans before they are activated and the harmfulness they would introduce to the host models when they are activated on real-life deep learning models, AlexNet and VGG16.
The results show the proposed PoTrojans has very low accident triggering rate and significant impact on their host models.

The shortcoming of the proposed PoTrojans is it could only be triggered by specific inputs, which limits its applicability.
We would continue working on the idea of neuron-level trojans and improve the triggering mechanism so that the PoTrojans could be triggered by inputs containing specific features.

\ifCLASSOPTIONcaptionsoff
  \newpage
\fi

\bibliographystyle{IEEEtran}
\bibliography{Mendeley.bib}{}

\begin{thebibliography}{10}
\providecommand{\url}[1]{#1}
\csname url@samestyle\endcsname
\providecommand{\newblock}{\relax}
\providecommand{\bibinfo}[2]{#2}
\providecommand{\BIBentrySTDinterwordspacing}{\spaceskip=0pt\relax}
\providecommand{\BIBentryALTinterwordstretchfactor}{4}
\providecommand{\BIBentryALTinterwordspacing}{\spaceskip=\fontdimen2\font plus
\BIBentryALTinterwordstretchfactor\fontdimen3\font minus
  \fontdimen4\font\relax}
\providecommand{\BIBforeignlanguage}[2]{{%
\expandafter\ifx\csname l@#1\endcsname\relax
\typeout{** WARNING: IEEEtran.bst: No hyphenation pattern has been}%
\typeout{** loaded for the language `#1'. Using the pattern for}%
\typeout{** the default language instead.}%
\else
\language=\csname l@#1\endcsname
\fi
#2}}
\providecommand{\BIBdecl}{\relax}
\BIBdecl

\bibitem{He2016}
X.~Z. S.~R. He, Kaiming and J.~Sun, ``{Deep residual learning for image
  steganalysis},'' in \emph{Proceedings of the IEEE conference on computer
  vision and pattern recognition}, 2016, pp. 770--778.

\bibitem{Silver2016}
\BIBentryALTinterwordspacing
D.~Silver, A.~Huang, C.~J. Maddison, A.~Guez, L.~Sifre, G.~Van Den~Driessche,
  J.~Schrittwieser, I.~Antonoglou, V.~Panneershelvam, M.~Lanctot, S.~Dieleman,
  D.~Grewe, J.~Nham, N.~Kalchbrenner, I.~Sutskever, T.~Lillicrap, M.~Leach,
  K.~Kavukcuoglu, T.~Graepel, and D.~Hassabis, ``{Mastering the game of Go with
  deep neural networks and tree search},'' \emph{Nature}, vol. 529, no. 7587,
  pp. 484--489, 2016. [Online]. Available:
  \url{http://dx.doi.org/10.1038/nature16961}
\BIBentrySTDinterwordspacing

\bibitem{Goodfellow2014a}
\BIBentryALTinterwordspacing
I.~J. Goodfellow, J.~Pouget-Abadie, M.~Mirza, B.~Xu, D.~Warde-Farley, S.~Ozair,
  A.~Courville, and Y.~Bengio, ``{Generative Adversarial Networks},'' pp. 1--9,
  2014. [Online]. Available: \url{http://arxiv.org/abs/1406.2661}
\BIBentrySTDinterwordspacing

\bibitem{Evtimov2017}
\BIBentryALTinterwordspacing
I.~Evtimov, K.~Eykholt, E.~Fernandes, T.~Kohno, B.~Li, A.~Prakash, A.~Rahmati,
  and D.~Song, ``{Robust Physical-World Attacks on Deep Learning Models},''
  2017. [Online]. Available: \url{http://arxiv.org/abs/1707.08945}
\BIBentrySTDinterwordspacing

\bibitem{Tehranipoor2009}
M.~Tehranipoor and F.~Koushanfar, ``{A Survey of Hardware Trojan Taxonomy and
  Detection},'' pp. 1--18, 2009.

\bibitem{Gu2017}
\BIBentryALTinterwordspacing
T.~Gu, B.~Dolan-Gavitt, and S.~Garg, ``{BadNets: Identifying Vulnerabilities in
  the Machine Learning Model Supply Chain},'' 2017. [Online]. Available:
  \url{http://arxiv.org/abs/1708.06733}
\BIBentrySTDinterwordspacing

\bibitem{Liu2017a}
\BIBentryALTinterwordspacing
Y.~Liu, S.~Ma, Y.~Aafer, W.-C. Lee, J.~Zhai, A.~Yingqi~Liu, W.~Wang, and
  X.~Zhang, ``{Trojaning Attack on Neural Networks},'' 2017. [Online].
  Available: \url{http://docs.lib.purdue.edu/cstech/1781}
\BIBentrySTDinterwordspacing

\bibitem{Geigel2013}
A.~Geigel, ``{Neural network Trojan},'' \emph{Journal of Computer Security},
  vol.~21, no.~2, pp. 191--232, 2013.

\bibitem{Krizhevsky2012}
A.~Krizhevsky, I.~Sutskever, and G.~E. Hinton, ``{ImageNet Classification with
  Deep Convolutional Neural Networks},'' \emph{Advances In Neural Information
  Processing Systems}, pp. 1--9, 2012.

\bibitem{Simonyan2014}
\BIBentryALTinterwordspacing
K.~Simonyan and A.~Zisserman, ``{Very Deep Convolutional Networks for
  Large-Scale Image Recognition},'' \emph{arXiv preprint arXiv:1409.1556}, pp.
  1--14, 9 2014. [Online]. Available: \url{http://arxiv.org/abs/1409.1556}
\BIBentrySTDinterwordspacing

\bibitem{Russakovsky2015}
O.~Russakovsky, J.~Deng, H.~Su, J.~Krause, S.~Satheesh, S.~Ma, Z.~Huang,
  A.~Karpathy, A.~Khosla, M.~Bernstein, A.~C. Berg, and L.~Fei-Fei, ``{ImageNet
  Large Scale Visual Recognition Challenge},'' \emph{International Journal of
  Computer Vision}, vol. 115, no.~3, pp. 211--252, 2015.

\bibitem{MichaelGuerzhoy}
\BIBentryALTinterwordspacing
M.~Guerzhoy, ``{AlexNet implementation + weights in TensorFlow}.'' [Online].
  Available: \url{http://www.cs.toronto.edu/~guerzhoy/tf_alexnet/}
\BIBentrySTDinterwordspacing

\bibitem{Frossard}
\BIBentryALTinterwordspacing
D.~Frossard, ``{VGG in TensorFlow}.'' [Online]. Available:
  \url{http://www.cs.toronto.edu/~frossard/post/vgg16/}
\BIBentrySTDinterwordspacing

\end{thebibliography}

\begin{IEEEbiography}[{\includegraphics[width=1in,height=1.25in,clip,keepaspectratio]{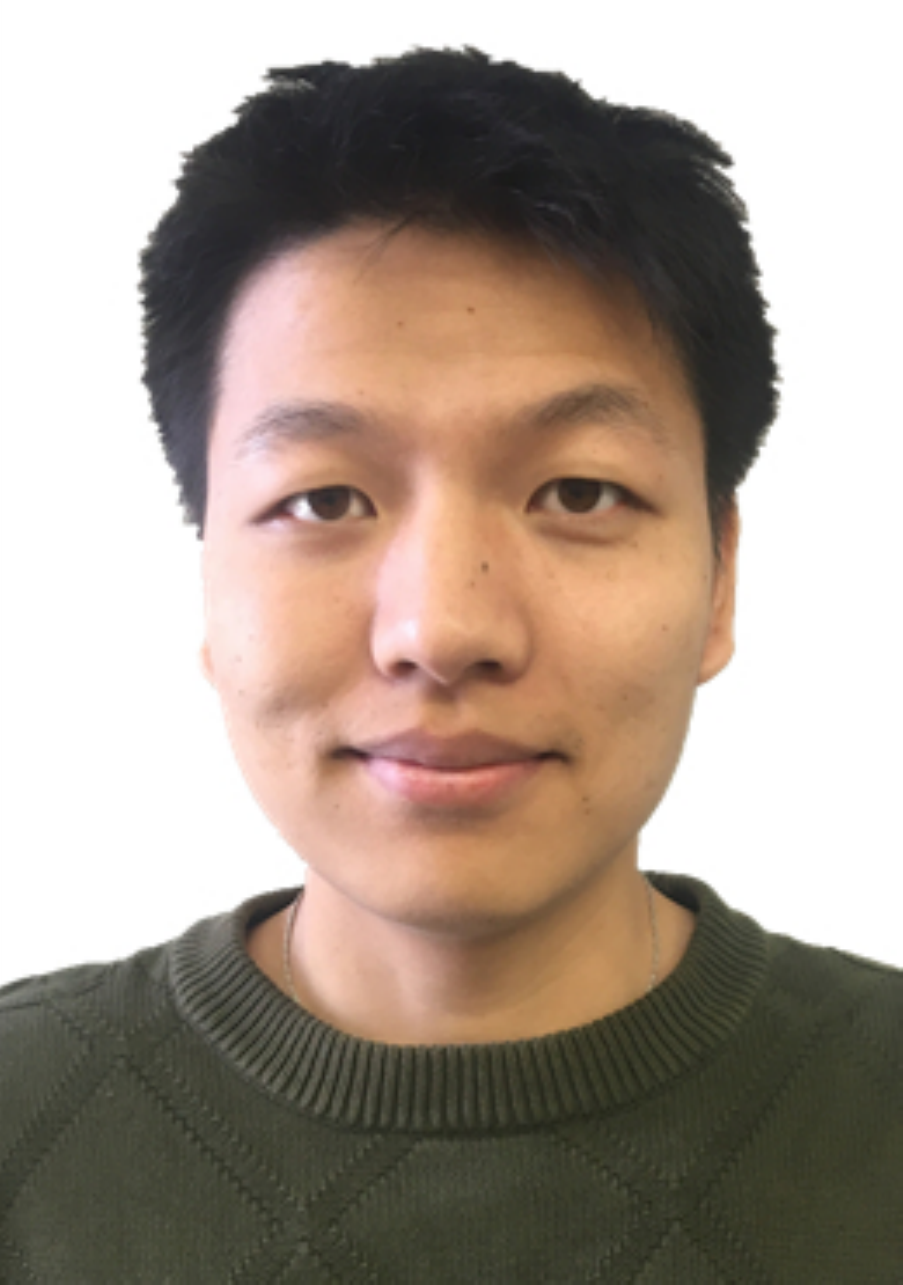}}]
{Minhui Zou} received the B.S. degree in computer science and technology from Chongqing University, China, in 2013.
Currently he is a Ph.D. student majoring in computer science and technology of the College of Computer Science, Chongqing University.
His current research interests include hardware security, IoT security and neural network security.
\end{IEEEbiography}

\begin{IEEEbiography}
[{\includegraphics[width=1in,height=1.25in,clip,keepaspectratio]{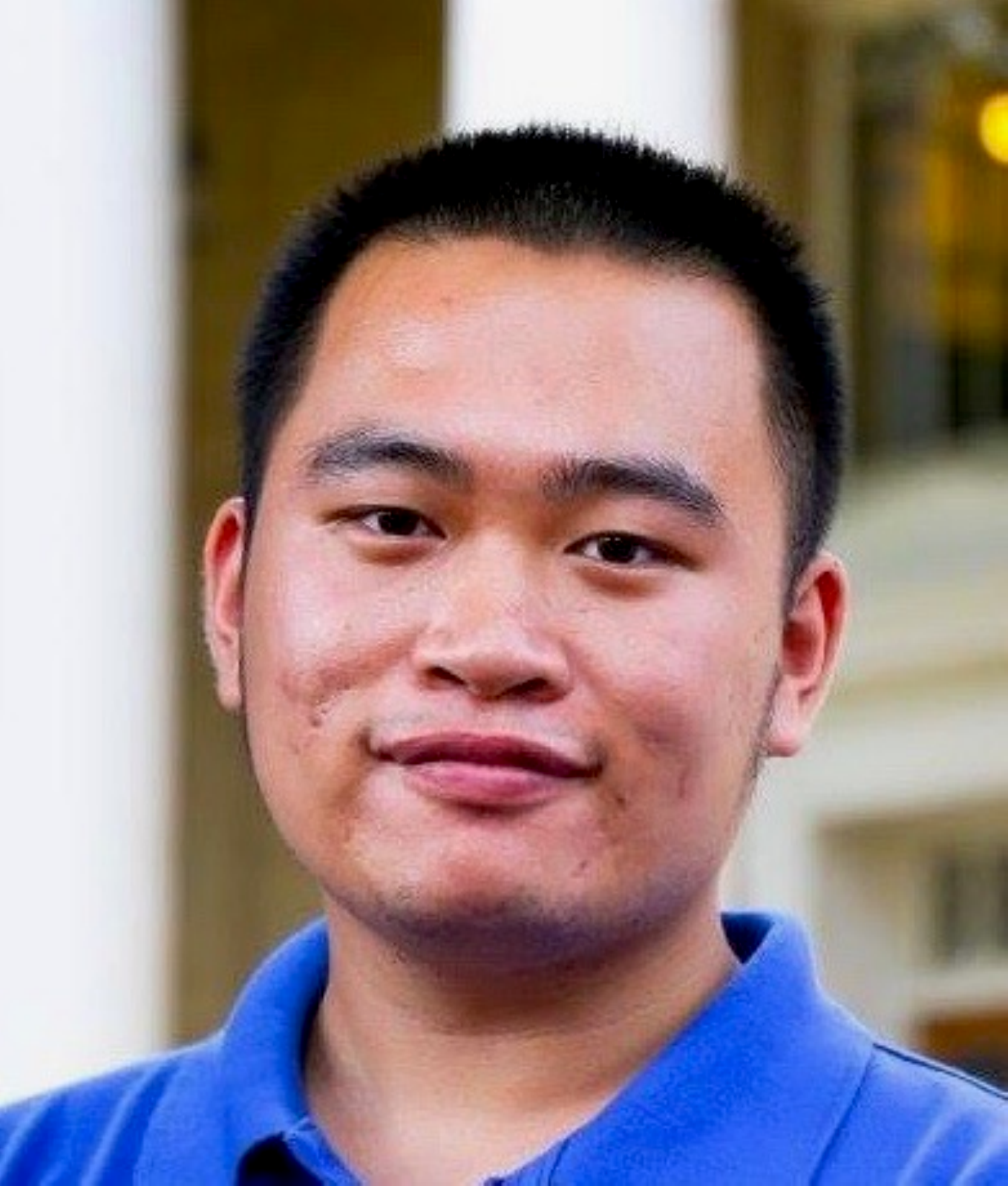}}]
{Yang~Shi} is a Ph.D. student in the Department of Computer Science, University of Georgia. 
He received his B.Eng degree in automation from Central South University, China, in 2015 and M.S. degree in computer science from University of Georgia in 2017. 
His research interests include distributed computing, machine learning, and Internet of Things (IoT).
\end{IEEEbiography}

\begin{IEEEbiography}[{\includegraphics[width=1in,height=1.25in,clip,keepaspectratio]{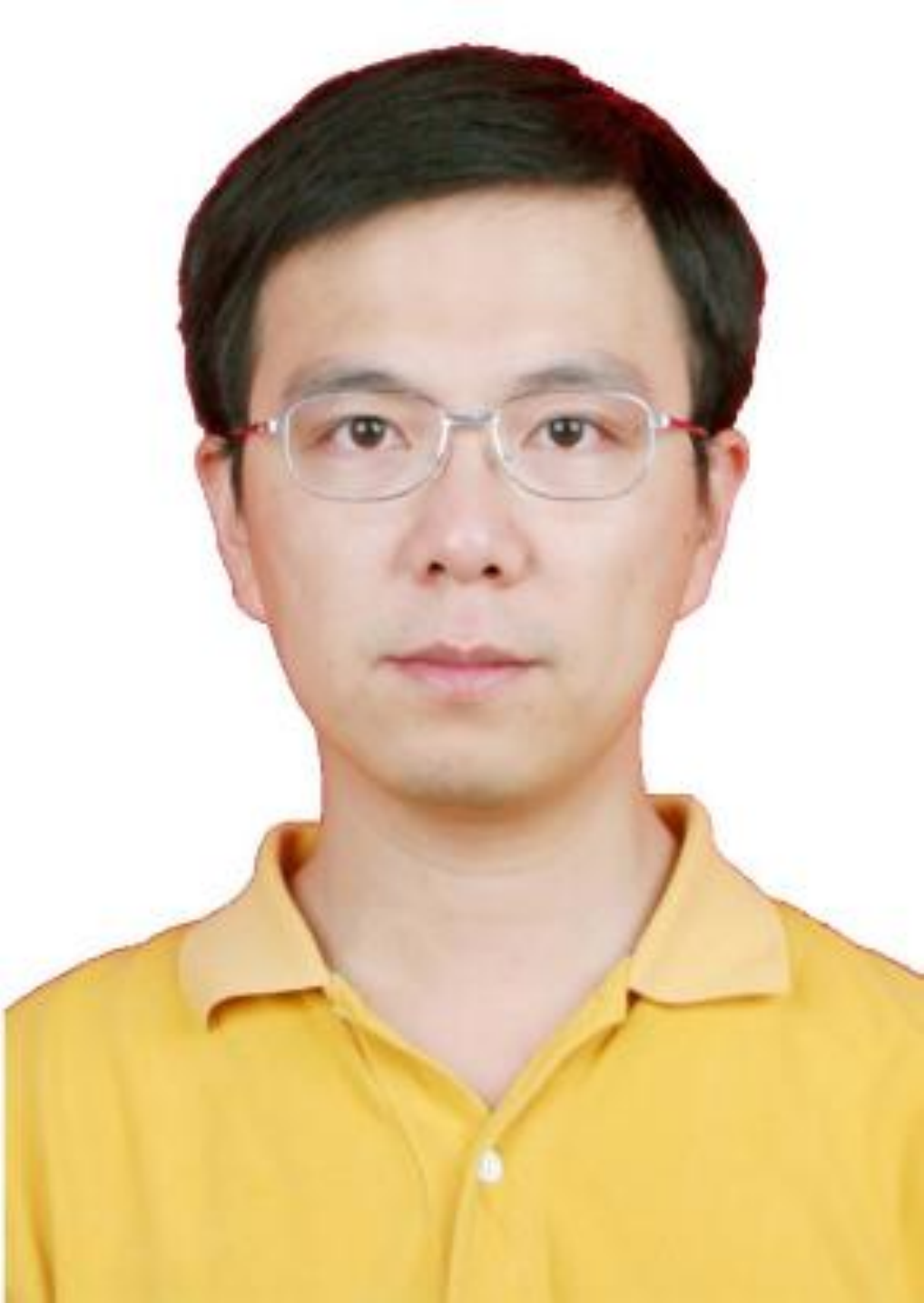}}]
{Chengliang~Wang} received his B.S. degree in mechatronics in 1996, the M.S. degree in precision instruments and machinery in 1999, and the Ph.D. degree in control theory and engineering in 2004, all from Chongqing University, China.
He is now a professor of Chongqing University.
His research interests include smart control for complex system, the theory and application of artificial intelligence, wireless network and RFID research.
He is a senior member of China computer science association and member of America Association of Computing Machinery.
\end{IEEEbiography}

\begin{IEEEbiography}
[{\includegraphics[width=1in,height=1.25in,clip,keepaspectratio]{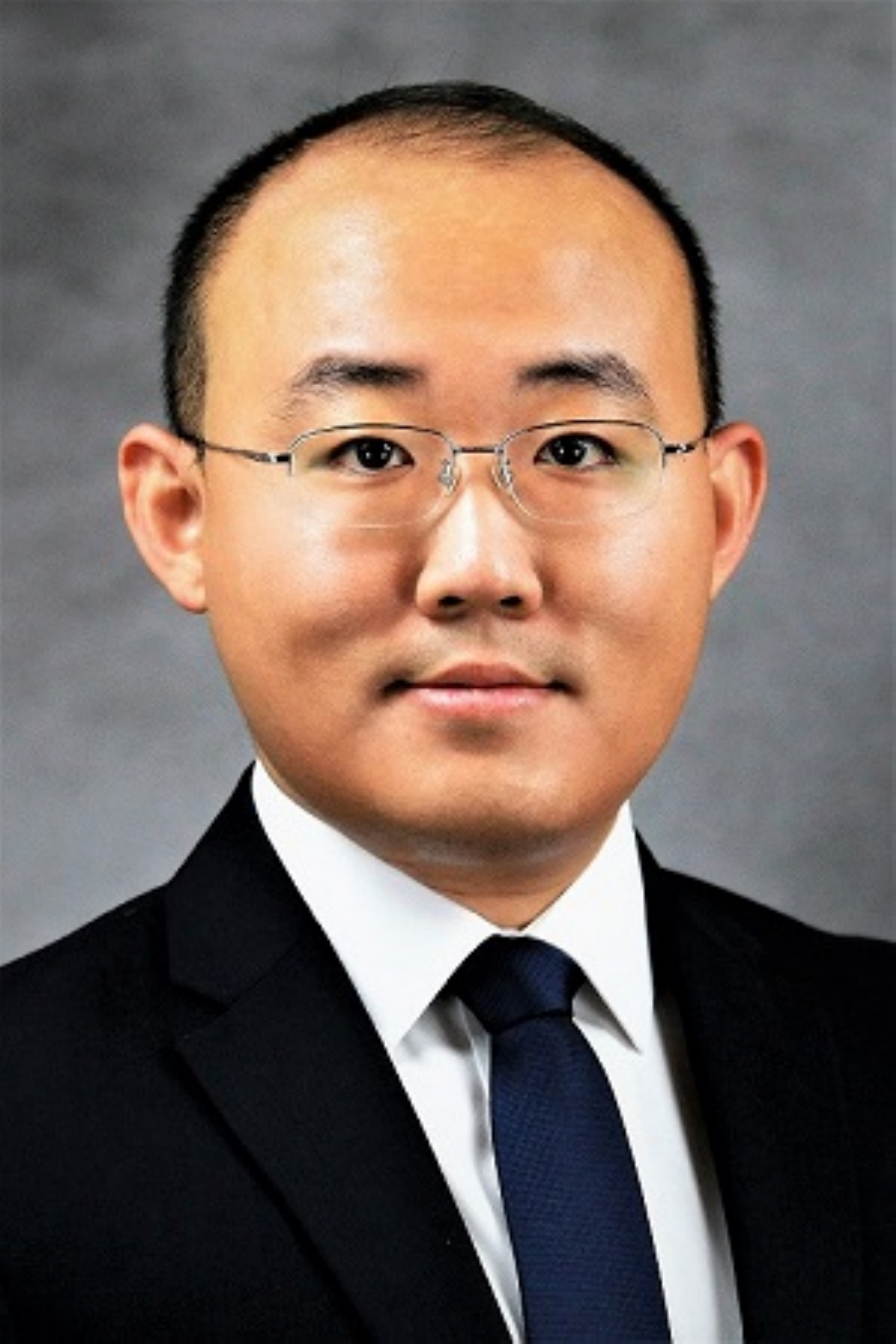}}]
{Fangyu Li} is a postdoctoral research associate in the College of Engineering, University of Georgia. He received his PhD in Geophysics from University of Oklahoma in 2017. His Master and Bachelor degrees are both in Electrical Engineering, obtained from Tsinghua University and Baihang University, respectively. His research interests include signal processing, seismic imaging, geophysical interpretation, machine learning and distributed system.
\end{IEEEbiography}

\begin{IEEEbiography}[{\includegraphics[width=1in,height=1.25in,clip,keepaspectratio]{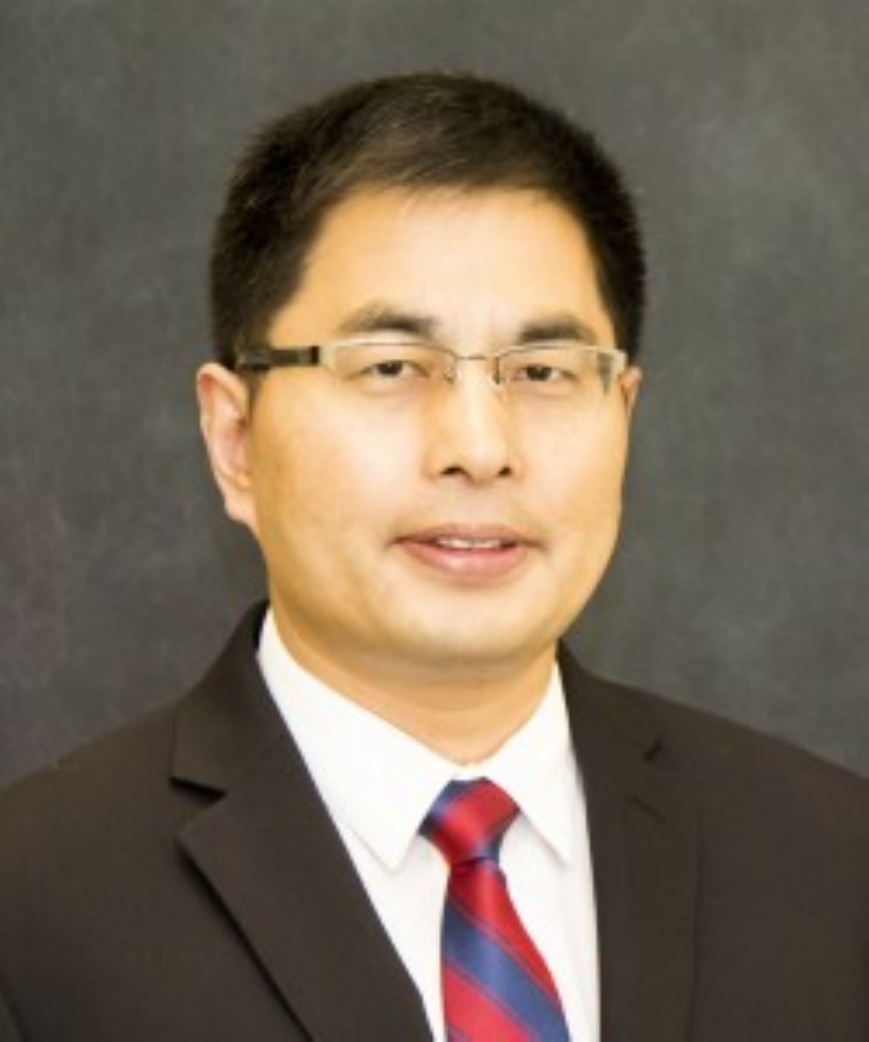}}]
{WenZhan~Song} is now Georgia Power Mickey A. Brown Professor in College of Engineering, University of Georgia. His research mainly focuses on sensor web, smart grid and smart environment where sensing, computing, communication and control play a critical role and need a transformative study. His research has received 6 million+ research funding from NSF, NASA, USGS, Boeing and etc since 2005.
\end{IEEEbiography}

\begin{IEEEbiography}[{\includegraphics[width=1in,height=1.25in,clip,keepaspectratio]{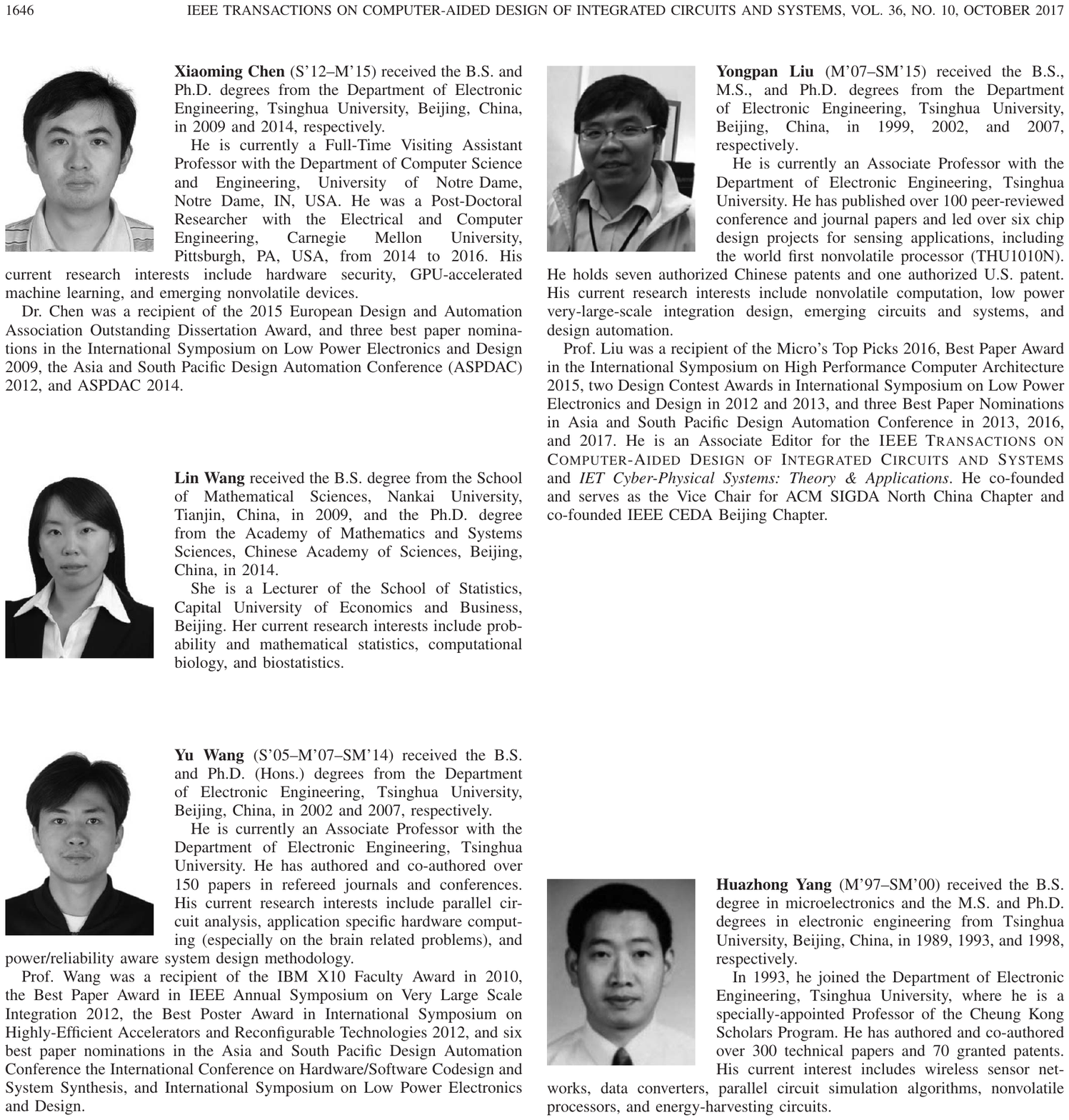}}]
{Yu~Wang} received the B.S. and Ph.D. (Hons.) degrees from the Department of Electronic Engineering, Tsinghua University, Beijing, China, in 2002 and 2007, respectively. 
He is currently an Associate Professor with the Department of Electronic Engineering, Tsinghua University. He has authored and co-authored over 150 papers in refereed  ournals and conferences.
His current research interests include parallel circuit analysis, application specific hardware computing (especially on the brain related problems), and power/reliability aware system design methodology.
Prof. Wang was a recipient of the IBM X10 Faculty Award in 2010, the Best Paper Award in IEEE Annual Symposium on Very Large Scale Integration 2012, the Best Poster Award in International Symposium on Highly-Efficient Accelerators and Reconfigurable Technologies 2012, and six best paper nominations in the Asia and South Pacific Design Automation
Conference the International Conference on Hardware/Software Codesign and System Synthesis, and International Symposium on Low Power Electronics and Design.
\end{IEEEbiography}

\end{document}